\documentstyle[epsf]{mn}

\input ols.sty

\title[Old Globular Clusters in the LMC]{HST Colour-Magnitude Diagrams of Six Old Globular Clusters in the LMC\thanks{Based on observations made with the NASA/ESA {\it Hubble Space Telescope}, obtained at the Space Telescope Science Institute, which is operated by the Association of Universities for Research in Astronomy, Inc. under NASA contract NAS5-26555.}}
\author[K.A.G. Olsen et al.]{K.A.G.~Olsen,$^1$ P.W.~Hodge,$^1$ M.~Mateo,$^2$ E.W.~Olszewski,$^3$ 
\newauthor
R.A.~Schommer,$^4$ N.B.~Suntzeff,$^4$ and A.R.~Walker$^4$\\
$^1$University of Washington, Department of Astronomy, Seattle, WA 98195, USA\\
$^2$University of Michigan, Department of Astronomy, Ann Arbor, MI 48109, USA\\
$^3$University of Arizona, Steward Observatory, Tucson, AZ 85721, USA\\
$^4$Cerro Tololo Inter-American Observatory, La Serena, Chile}

\begin{document}

\maketitle

\begin{abstract}
We report on HST observations of six candidate old globular clusters in the
Large 
Magellanic Cloud: NGC 1754, NGC 1835, NGC 1898,  NGC 1916, NGC 2005 and NGC 
2019. Deep exposures with the F555W and F814W filters provide us with colour-magnitude
diagrams that 
reach to an apparent magnitude in $V$ of $\sim$25, well below the main 
sequence turnoff. These particular clusters are involved with
significantly high LMC  
field star densities and care was taken to subtract the field stars from
the cluster colour-magnitude diagrams 
accurately. In two cases there is significant variable reddening across at least
part of the image, 
but only for NGC 1916 does the differential reddening preclude accurate
measurements 
of the CMD characteristics. The morphologies of the colour-magnitude
diagrams match 
well those of Galactic globular clusters of similar metallicity. All six
have well-developed 
horizontal branches, while four clearly have stars on both sides of the RR Lyrae gap.    The abundances obtained from measurements of the height of the red giant branch above the level of the horizontal branch are 0.3 dex higher, on average, than previously measured spectroscopic abundances.  Detailed
comparisons 
with Galactic globular cluster fiducials show that all six clusters are old
objects, very similar in 
age to classical Galactic globulars such as M5, with little age spread among the clusters.  This result is consistent with ages derived by measuring the magnitude difference between the horizontal branch and main sequence turnoff.  We also find a similar chronology by comparing the horizontal branch morphologies and abundances with the horizontal branch evolutionary tracks of Lee, Demarque, \& Zinn (1994).  Our results imply that the LMC formed at the same time as the Milky Way Galaxy.
\end{abstract}

\section{Introduction}
	It has long been known that the LMC contains several clusters that strongly 
resemble normal globular clusters of the Galaxy (Shapley 1930). Several of
these objects, 
however, have turned out to be of intermediate or even young age, in spite
of their 
outward appearance of large size and luminosity. Some years ago one of us, in an attempt to use crude colour-magnitude diagrams, selected
35 clusters 
that seemed to be old globular clusters (Hodge 1960). Truly old
clusters are 
actually rarer, however; a recent review of the old populations of the MCs
lists only 14 
objects that are candidates for clusters that might be as old as the oldest
Galactic 
globulars (Olszewski, Suntzeff and Mateo 1996; see also Westerlund 1997).
However, 
precise ages are very difficult to determine from the ground, as the main
sequence turnoff (hereafter MSTO) occurs at a $V$ of approximately 23, too faint for
accurate ground-based 
photometry in crowded regions of the LMC. For the less crowded
outlying
clusters NGC 1466, NGC 1786, NGC 1841, NGC 2210, NGC 2257, and Reticulum, ground-based CMDs approaching and going below the MSTO have
demonstrated 
that they are quite old (Brocato et al. 1996; Walker 1990, 1992a, 1992b), with NGC 2257 perhaps being $\sim$2 Gyr younger than the bulk of Milky Way clusters (Testa et al. 1995).  With uncertainties in age of typically $\ga$3 Gyr, however, the data are not 
accurate enough to fully determine whether the clusters are truly as old as the
classical Galactic 
globulars or are 2-3 gigayears younger, as has been suggested on the
basis of their 
horizontal branch (hereafter HB) morphologies (Da Costa 1993).  This question is of strong interest both for establishing the timescale of formation of the LMC and for testing age as the candidate second parameter affecting HB morphology, a topic of currently hot debate (cf. Lee 1992; Lee, Demarque, \& Zinn 1994, hereafter LDZ; Stetson, VandenBerg, \& Bolte 1996; Sarajedini, Chaboyer, \& Demarque 1997).

This paper reports on a {\it Hubble Space Telescope} program to determine
colour-magnitude 
diagrams for the six inner globular clusters of the LMC, for which
ground-based data 
would be especially difficult to interpret because of the photometric
effects of crowding.  
The clusters and their characteristics (taken from Olszewski et al. 1996)
are listed in 
Table 1. These clusters, which presumably formed deep in the LMC gravitational potential well, are excellent probes of the early formation of the main mass of the LMC.  In conjunction with the HST study of Hodge 11 by Mighell et al. (1996) and the comprehensive study with HST of the outlying old LMC clusters currently underway (Johnson \& Bolte 1997), our results will be useful for testing models of the formation of the LMC, much as observations of globular clusters in the Milky Way have evaluated formation models of the Galaxy.  In the following we describe the observations in Section 2, detail
our reductions 
in Section 3, discuss field star decontamination in Section 4, and derive
the colour-magnitude diagrams (CMDs) for the clusters in Section 5. In Section 6, we present the ages of the clusters implied by three age-dating techniques: comparison of our CMDs with Milky Way globular cluster fiducial sequences through the ``horizontal method" (VandenBerg, Bolte, \& Stetson 1990, hereafter VBS), measurement of $\Delta V{\rm ^{TO}_{HB}}$ (``vertical method"), and comparison of our HB data with HB evolutionary models from LDZ.  These ages are dependent on the adopted abundances, which we measure from the CMDs and compare with previously measured spectroscopic abundances (Olszewski et al. 1991, herafter O91).  We also describe in Section 6 our estimates of the reddenings and distances of the clusters.  These estimates are not direct measurements, but rely on knowledge of the reddenings and distances to Galactic globular clusters and on an assumed $M_V$(RR)-[Fe/H] relationship.

\begin{table}
 \caption{Basic LMC Globular Cluster Parameters}
 \begin{tabular}{@{}lcccc@{}}
Cluster & RA (2000.0) & Dec & $V$ (int.) & n(RR) \\
\hline
NGC 1754\dotfill & 04h55m & -70\fdg31\arcmin & 11.4 & $-$\\
NGC 1835\dotfill & 05h05m & -69\fdg28\arcmin & 9.5 & 35 \\
NGC 1898\dotfill & 05h17m & -69\fdg43\arcmin & 11.1 & $-$ \\
NGC 1916\dotfill & 05h19m & -69\fdg27\arcmin & 9.9 & $-$ \\
NGC 2005\dotfill & 05h30m & -69\fdg45\arcmin & 11.2 & $-$ \\
NGC 2019\dotfill & 05h32m & -70\fdg12\arcmin & 10.7 & 0 \\
\hline
\end{tabular}
\end{table}

\section{Observations}
Observations of the clusters were taken during Cycle 5 of HST.  Each cluster was observed for the duration of one orbit of the spacecraft.  Table 2 contains a summary of the observation log.  For these observations, we used the low gain setting of the WFPC2 camera (GAIN=7).  Multiple exposures were taken through each filter to aid in cosmic ray removal, but no dithering technique was used.  We took both long and short exposures to provide unsaturated photometry of as many stars as possible.  The long exposures reach limiting magnitudes of $\sim$24.5 in F555W and $\sim$23.5 in F814W, or $\sim$1.5 magnitudes below the main sequence turnoff in $V$.  The images were processed through the standard STScI reduction pipeline prior to our receipt of them.

In order to check the zero point of the HST photometry, one of us (A.W.) obtained images of each cluster with the CTIO 1.5-m and Tek 2048 CCD.  The field of view of the camera is 8\farcm3, or $\sim$3 times the size of the WFPC2 field of view.  Images were taken through CTIO copies of the {\it HST} F555W and F814W
filters on the nights of January 23-26, 1995 under photometric conditions and in good seeing (FWHM$\sim$1\arcsec).  Standard stars from Landolt (1992) and from the $\omega$ Cen field used to calibrate WFPC2 photometry were observed.  Careful photometry using DAOPHOT/ALLSTAR was performed on suitably isolated stars in the ground-based frames and the photometry transformed to Johnson/Kron-Cousins $V$ and $I$.  The comparison of the photometry from these ground-based images with the HST photometry is described in section 3.6.

\begin{table*}
\caption{Summary of Observing Log}
\begin{tabular}{@{}llccc@{}}
Cluster & 
Files &
Filter &
Exp. time (sec.) &
Date (DD/MM/YY) \\
\hline
NGC 1754\dotfill & u2xq0101t\_cvt.$\ast$, u2xq0102t\_cvt.$\ast$ & F555W & 20 & 21/10/95 \\
& u2xq0103t\_cvt.$\ast$, u2xq0104t\_cvt.$\ast$, u2xq0105t\_cvt.$\ast$ & F555W & 500 & 21/10/95 \\
& u2xq0106t\_cvt.$\ast$, u2xq0107t\_cvt.$\ast$, u2xq0108t\_cvt.$\ast$ & F814W & 20 & 21/10/95 \\
& u2xq0109t\_cvt.$\ast$, u2xq010at\_cvt.$\ast$, u2xq010bt\_cvt.$\ast$ & F814W & 600 & 21/10/95 \\
NGC 1835\dotfill & u2xq0201t\_cvt.$\ast$, u2xq0202t\_cvt.$\ast$ & F555W & 20 & 18/10/95 \\
& u2xq0203t\_cvt.$\ast$, u2xq0204t\_cvt.$\ast$, u2xq0205t\_cvt.$\ast$ & F555W & 500 & 18/10/95 \\
& u2xq0206t\_cvt.$\ast$, u2xq0207t\_cvt.$\ast$, u2xq0208t\_cvt.$\ast$ & F814W & 20 & 18/10/95 \\
& u2xq0209t\_cvt.$\ast$, u2xq020at\_cvt.$\ast$, u2xq020bt\_cvt.$\ast$ & F814W & 600 & 18/10/95 \\
NGC 1898\dotfill & u2xq0301t\_cvt.$\ast$, u2xq0302t\_cvt.$\ast$ & F555W & 20 & 10/12/95 \\
& u2xq0303t\_cvt.$\ast$, u2xq0304t\_cvt.$\ast$, u2xq0305t\_cvt.$\ast$ & F555W & 500 & 10/12/95 \\
& u2xq0306t\_cvt.$\ast$, u2xq0307t\_cvt.$\ast$, u2xq0308t\_cvt.$\ast$ & F814W & 20 & 10/12/95 \\
& u2xq0309t\_cvt.$\ast$, u2xq030at\_cvt.$\ast$, u2xq030bt\_cvt.$\ast$ & F814W & 600 & 10/12/95 \\
NGC 1916\dotfill & u2xq0401t\_cvt.$\ast$, u2xq0402t\_cvt.$\ast$ & F555W & 20 & 10/12/95 \\
& u2xq0403t\_cvt.$\ast$, u2xq0404t\_cvt.$\ast$, u2xq0405t\_cvt.$\ast$ & F555W & 500 & 10/12/95 \\
& u2xq0406t\_cvt.$\ast$, u2xq0407t\_cvt.$\ast$, u2xq0408t\_cvt.$\ast$ & F814W & 20 & 10/12/95 \\
& u2xq0409t\_cvt.$\ast$, u2xq040at\_cvt.$\ast$, u2xq040bt\_cvt.$\ast$ & F814W & 600 & 10/12/95 \\
NGC 2005\dotfill & u2xq0501t\_cvt.$\ast$, u2xq0502t\_cvt.$\ast$ & F555W & 20 & 19/10/95 \\
& u2xq0503t\_cvt.$\ast$, u2xq0504t\_cvt.$\ast$, u2xq0505t\_cvt.$\ast$ & F555W & 500 & 19/10/95 \\
& u2xq0506t\_cvt.$\ast$, u2xq0507t\_cvt.$\ast$, u2xq0508t\_cvt.$\ast$ & F814W & 20 & 19/10/95 \\
& u2xq0509t\_cvt.$\ast$, u2xq050at\_cvt.$\ast$, u2xq050bt\_cvt.$\ast$ & F814W & 600 & 19/10/95 \\
NGC 2019\dotfill & u2xq0601t\_cvt.$\ast$, u2xq0602t\_cvt.$\ast$ & F555W & 20 & 18/10/95 \\
& u2xq0603t\_cvt.$\ast$, u2xq0604t\_cvt.$\ast$, u2xq0605t\_cvt.$\ast$ & F555W & 500 & 18/10/95 \\
& u2xq0606t\_cvt.$\ast$, u2xq0607t\_cvt.$\ast$, u2xq0608t\_cvt.$\ast$ & F814W & 20 & 18/10/95 \\
& u2xq0609t\_cvt.$\ast$, u2xq060at\_cvt.$\ast$, u2xq060bt\_cvt.$\ast$ & F814W & 600 & 18/10/95 \\
\hline
\end{tabular}
\end{table*}
\section{Reductions, Photometry, and Calibration of WFPC2 Images}

\begin{figure*}
\plotone{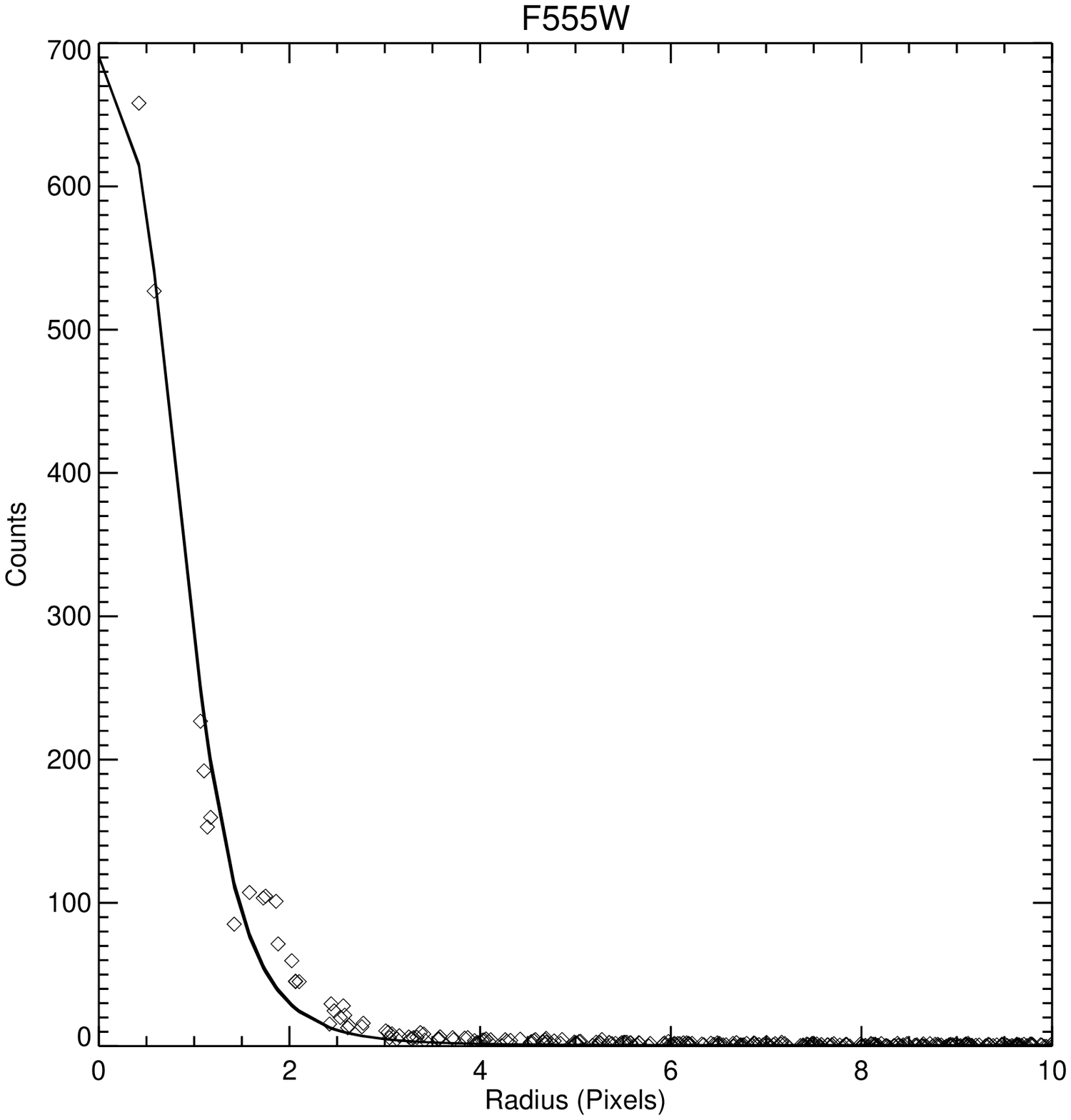}
\plotone{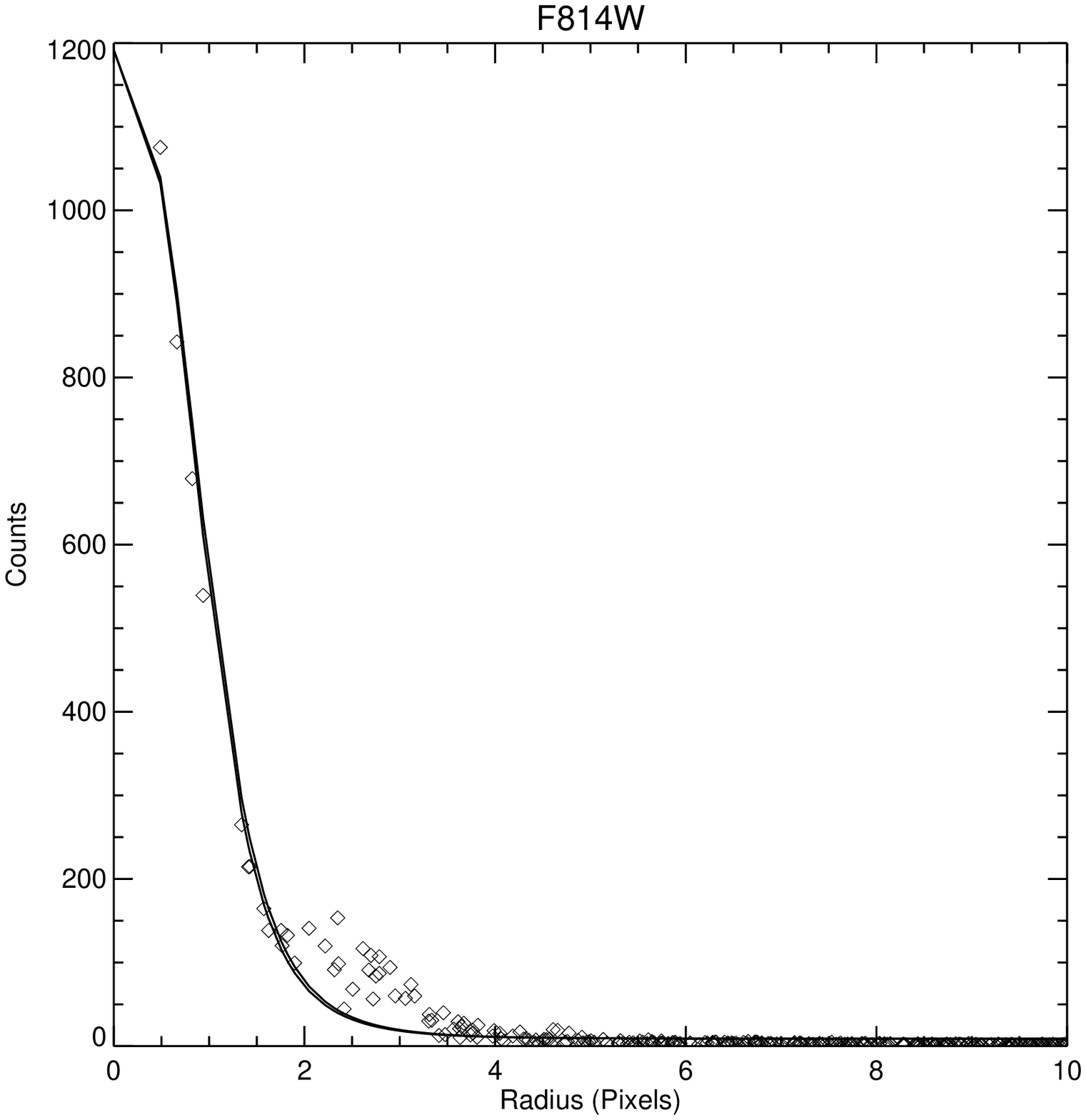}
\caption{Fit of the DoPHOT analytical PSF used in the photometry to a sample star from the WFPC2 frames.  While the PSF fits the core of the star and the background well, it doesn not fit the broad wings most prominent in the F814W images.}
\end{figure*}

\subsection{Image reduction and corrections}
Our first step was to check the image alignment for each set of exposures in
each filter.  In all cases, the alignment is better than a fraction of pixel.
We made no attempt to align the images further.

We removed the cosmic rays from the images with the IRAF\footnote{IRAF is written and supported by the IRAF programming group at the National Optical Astronomy Observatories (NOAO) in Tucson, Arizona.  NOAO is operated by the Association of Universities for Research in Astronomy, Inc. under cooperative agreement with the National Science Foundation.} STSDAS task {\it
crrej}.  From multiple equal-length exposures of the same field in each
filter, {\it crrej} produces a cosmic ray-free image by averaging the stack of
exposures after rejecting pixels with values too high compared to an initial
guess at the uncontanimated pixel values.  The threshold outside of which
pixels are rejected is set by the noise characteristics of the image and
through adjustable parameters.  Because there are fractional pixel offsets
between successive frames, the sharp cores of stars illuminate the pixels of the successive frames differently, and are often interpreted as
cosmic rays by the standard rejection procedure.  By setting the parameter
{\it scalenoise} to 10 per cent, which raises the rejection threshold by 10 per cent of the
pixel value, we insured that the cores of stars
were left unaltered while still removing many cosmic rays.

We accounted for CTE effects by using the corrections suggested by Holtzman et al. (1995a).  The backgrounds in the long exposures are in the 30-200 e$^-$ range,
while the short exposures have backgrounds of 2-10 e$^-$'s.  For the short
exposures, we multiplied an image containing a 4 per cent ramp into the data frames,
while we used a 2 per cent ramp for the long exposures.  Subsequent to our image
reduction, Whitmore \& Heyer (1997) published a report indicating that CTE
corrections should be a function of x- and y-position on the chip as well as
star counts and background counts, but the effects are small and we have not included them.

Because we are doing point source photometry, we need to correct for the fact
that the flat-fielding process adjusts the pixel counts so that surface
brightness is preserved with position on the chip.  We used the Holtzman et
al. (1995a) coordinate transformations to create pixel distortion maps for each
chip, which we multiplied into the data frames.

We used the DQF images to create bad pixel masks for each group of exposures in each filter.  These masks include macroscopic charge traps, bad columns, questionable pixels, and pixels saturated by stars in the field.  By setting the masked pixels in the data frames 
to a low value, they are ignored during photometry.

\begin{figure*}
\plotone{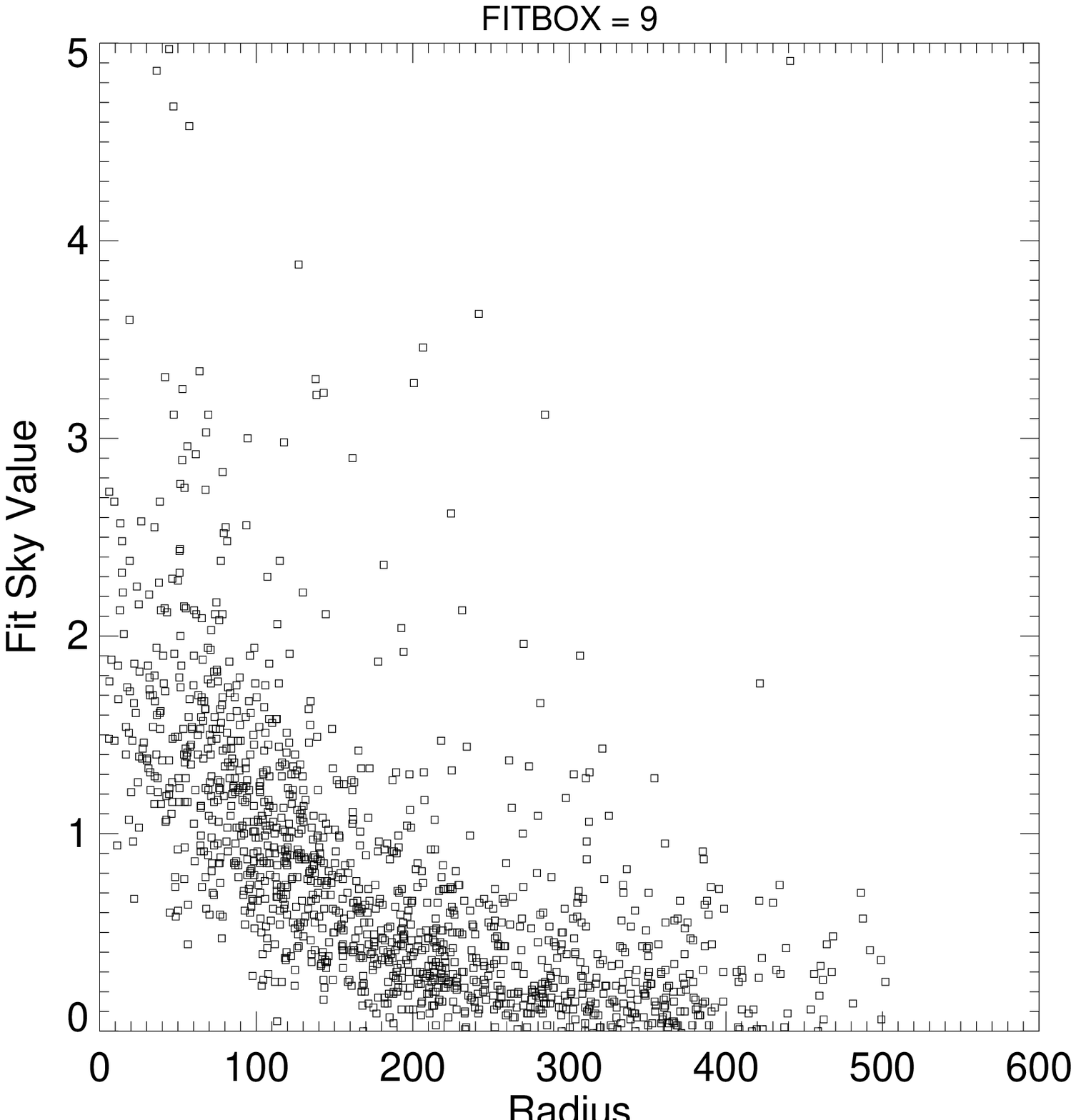}
\plotone{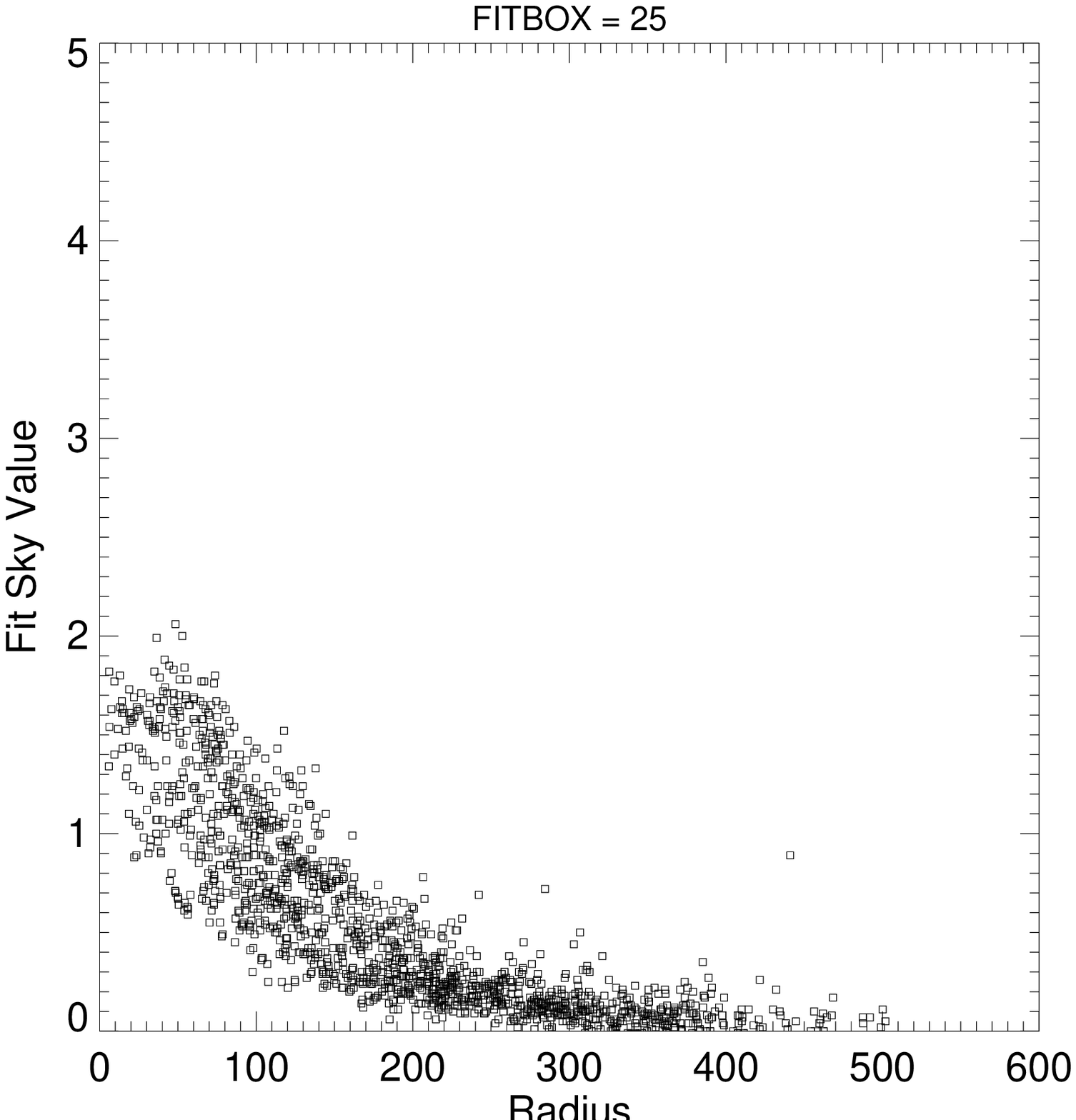}
\\
\hspace{0.8cm}
\plotone{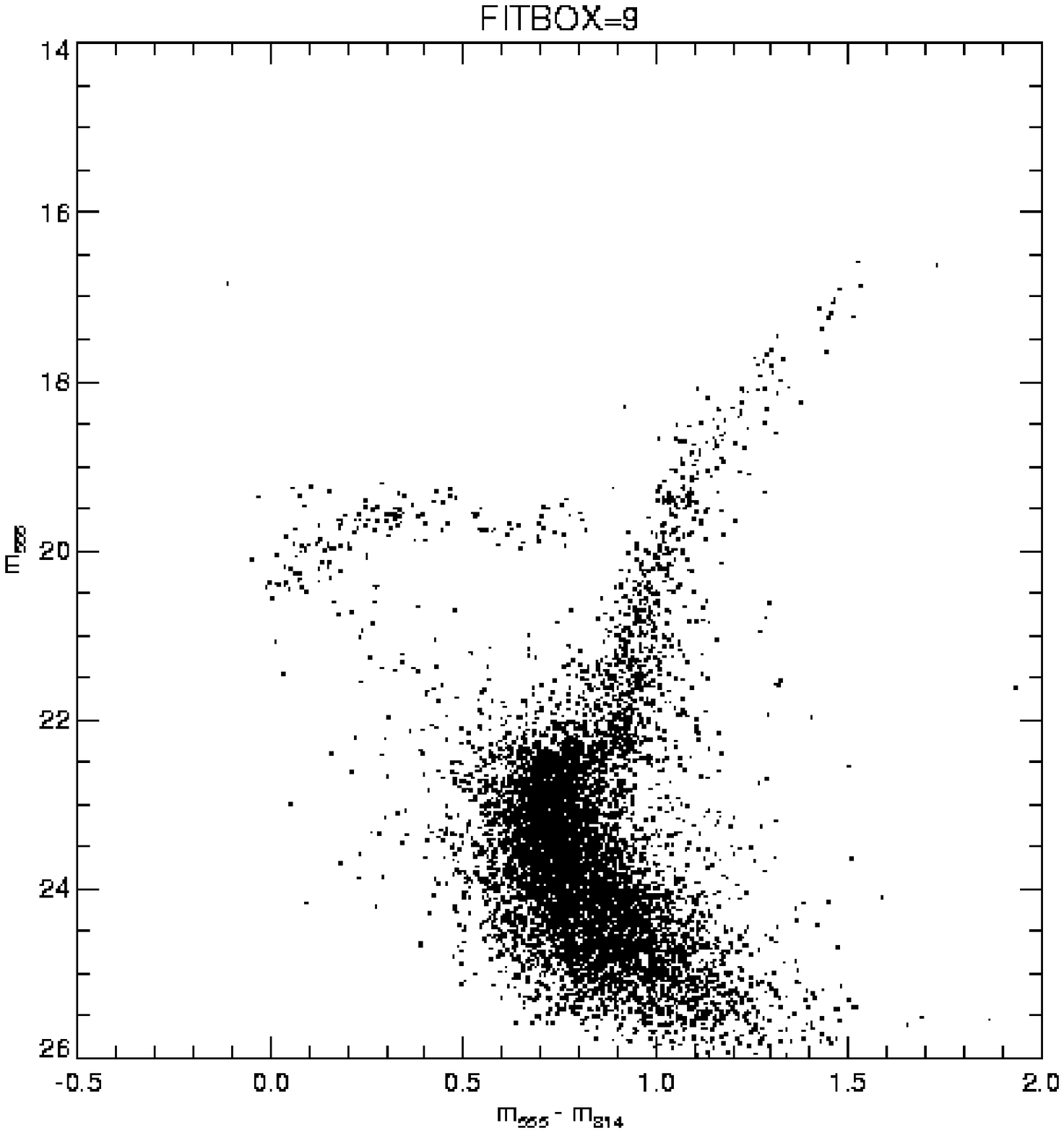}
\plotone{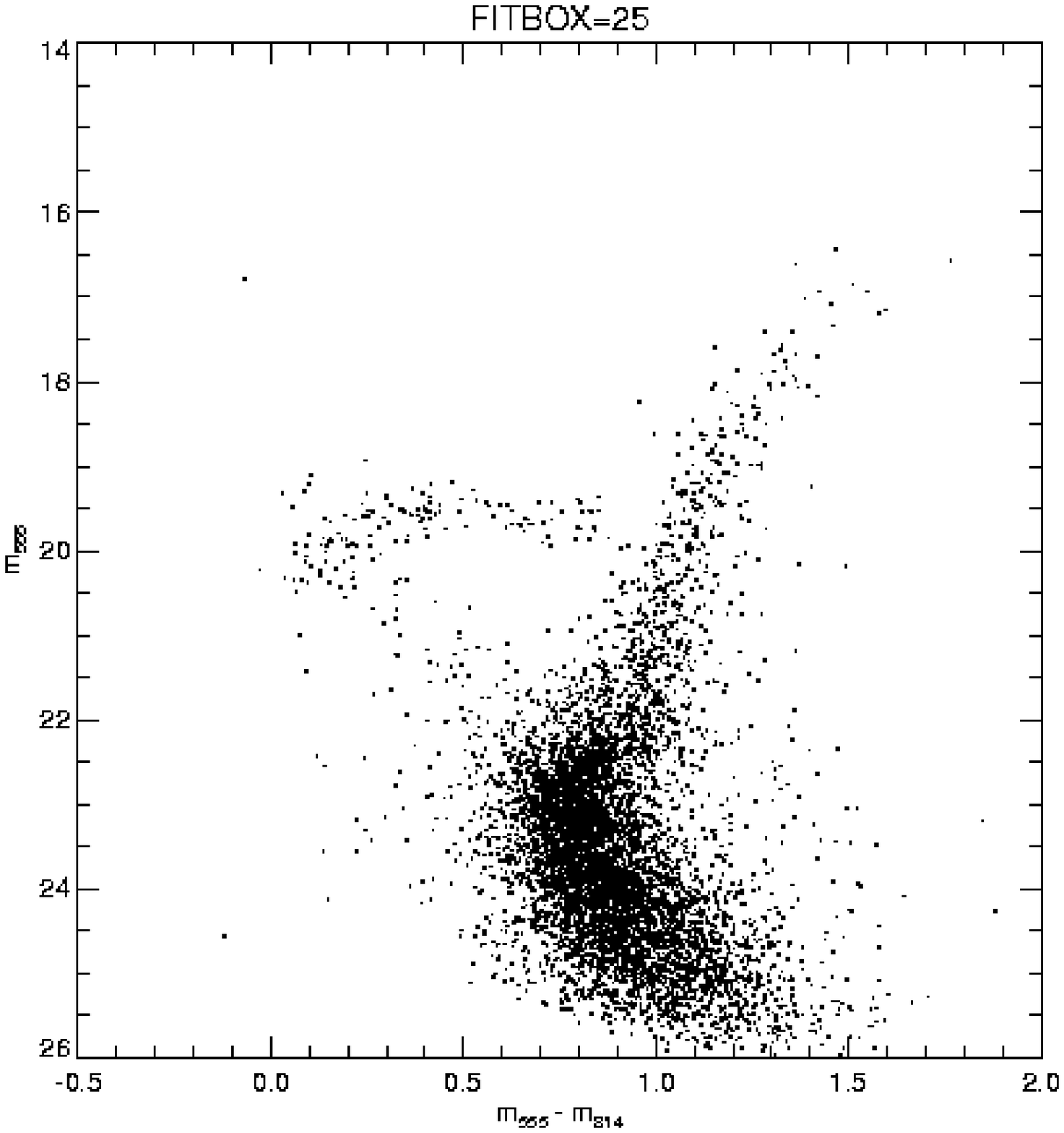}
\caption{Plots showing the effect of different sizes of the fitting box used in the DoPHOT photometry.  While using a larger fitting box produces better background fits, crowding limits the accuracy of the photometry.}
\end{figure*}

\subsection{DoPHOT Photometry}
In order to achieve the best photometry for the faint stars, which are crucial for determining cluster ages, it is necessary to resort to profile-fitting rather than aperture photometry.  For its speed and ease of use, we chose to use the program DoPHOT (Schechter, Mateo, \& Saha 1993), version 2.5, with modifications performed by Eric Deutsch
to handle our floating point images.  DoPHOT's operation is controlled by a file containing user-adjustable parameters, the most important of which are described in detail below.  

We chose to use DoPHOT's median filter to generate the model background above which DoPHOT identifies objects.  We found that DoPHOT's other background options, the plane and Hubble models, did not adequately fit the background, becoming weighted too heavily by the bright cluster cores and causing stars at the outskirts of the clusters to escape detection.  We used a minimum threshold for object detection of 2$\sigma$ above the background.

We adjusted the coefficients of DoPHOT's power law point-spread function (PSF) by simultaneously inspecting the fit of the PSF to a small sample of stellar radial profiles and by minimizing the aperture correction, described in Section 3.3.  We fit the coefficients independently for each filter and PC or WF chip.  Table 3 contains the final adopted values of $\beta_4$ and $\beta_6$, and Fig. 1 shows a sample fit of the PSF to a stellar radial profile.  Although the PSF fits well in the core of the star, the wing shows structure that is not fit by the PSF, most noticeably in the F814W filter.  Hence, the PSF-subtracted image shows halos around the positions of subtracted stars.  In order to prevent DoPHOT from identifying too many spurious objects in these residuals, we added extra noise to DoPHOT's noise array.

\begin{figure*}
\plotone{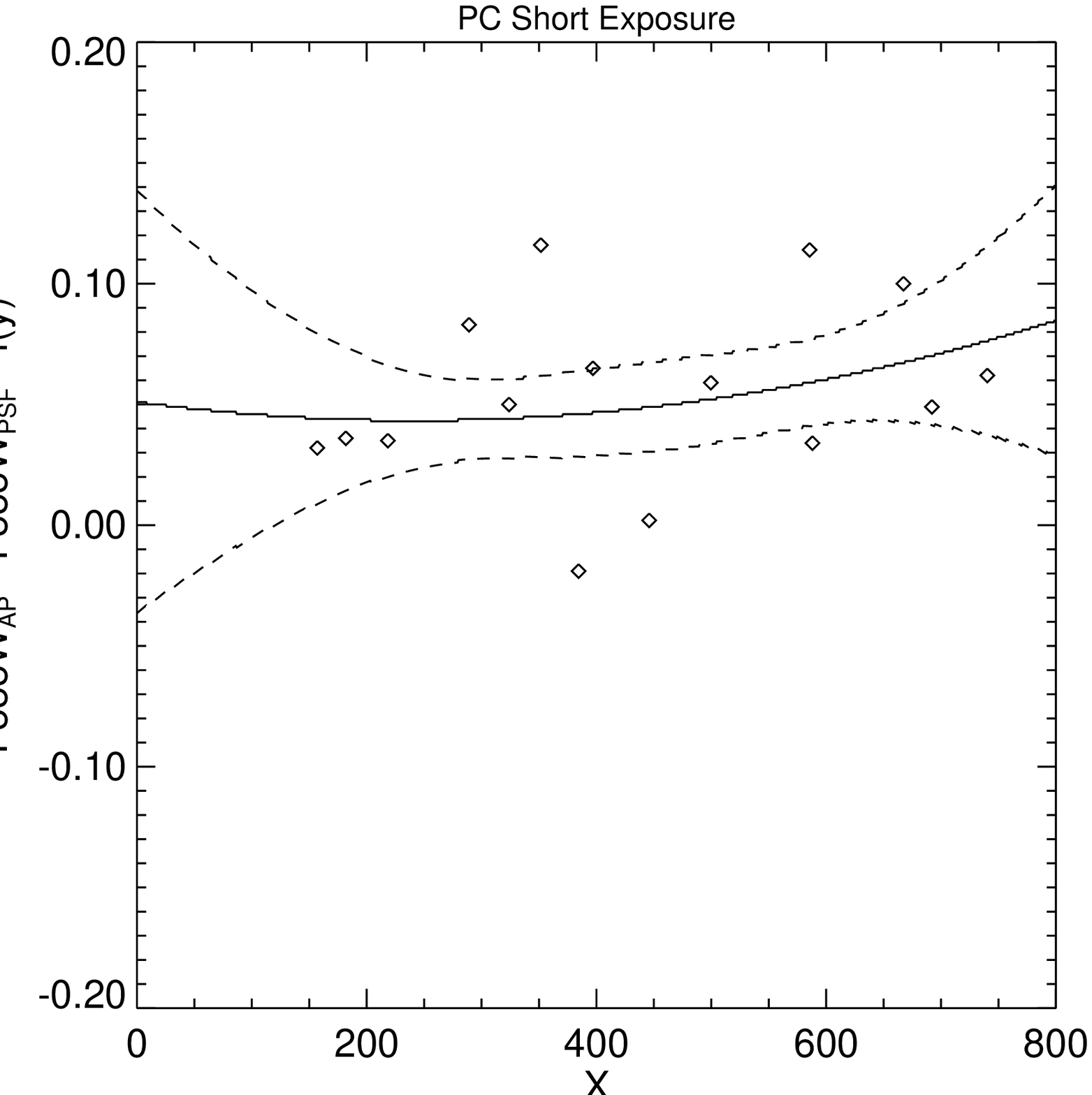}
\plotone{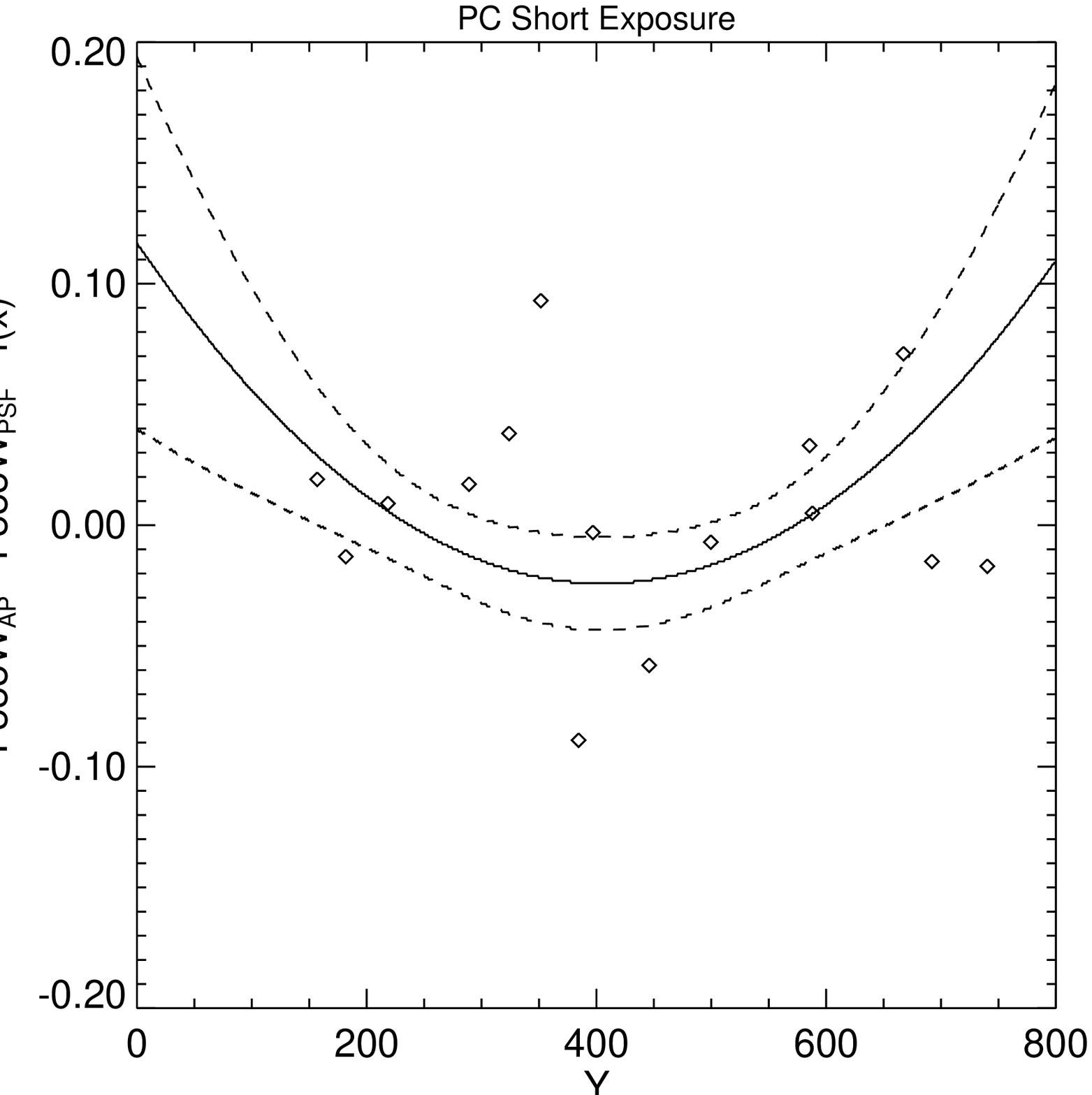}
\\
\plotone{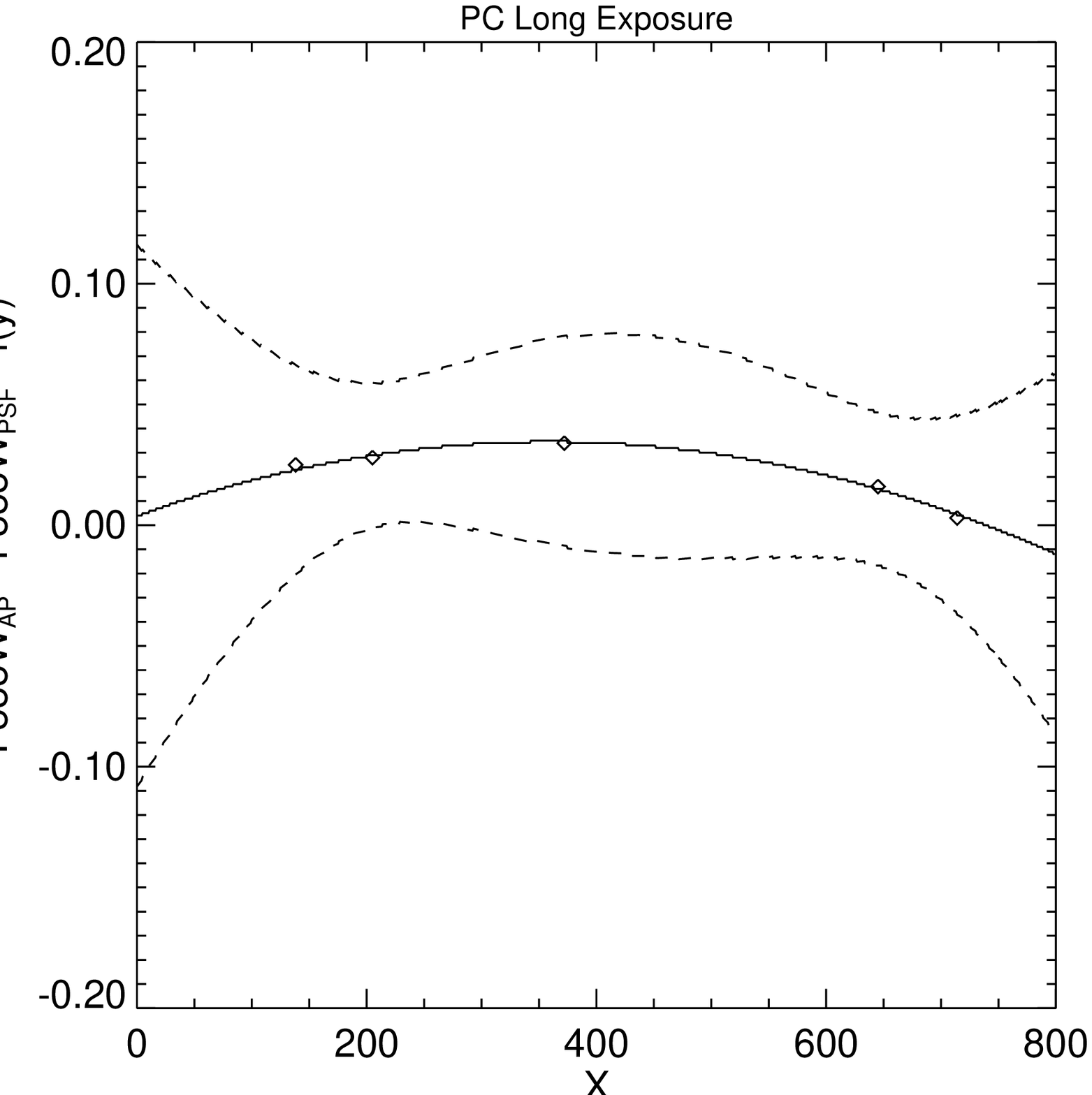}
\plotone{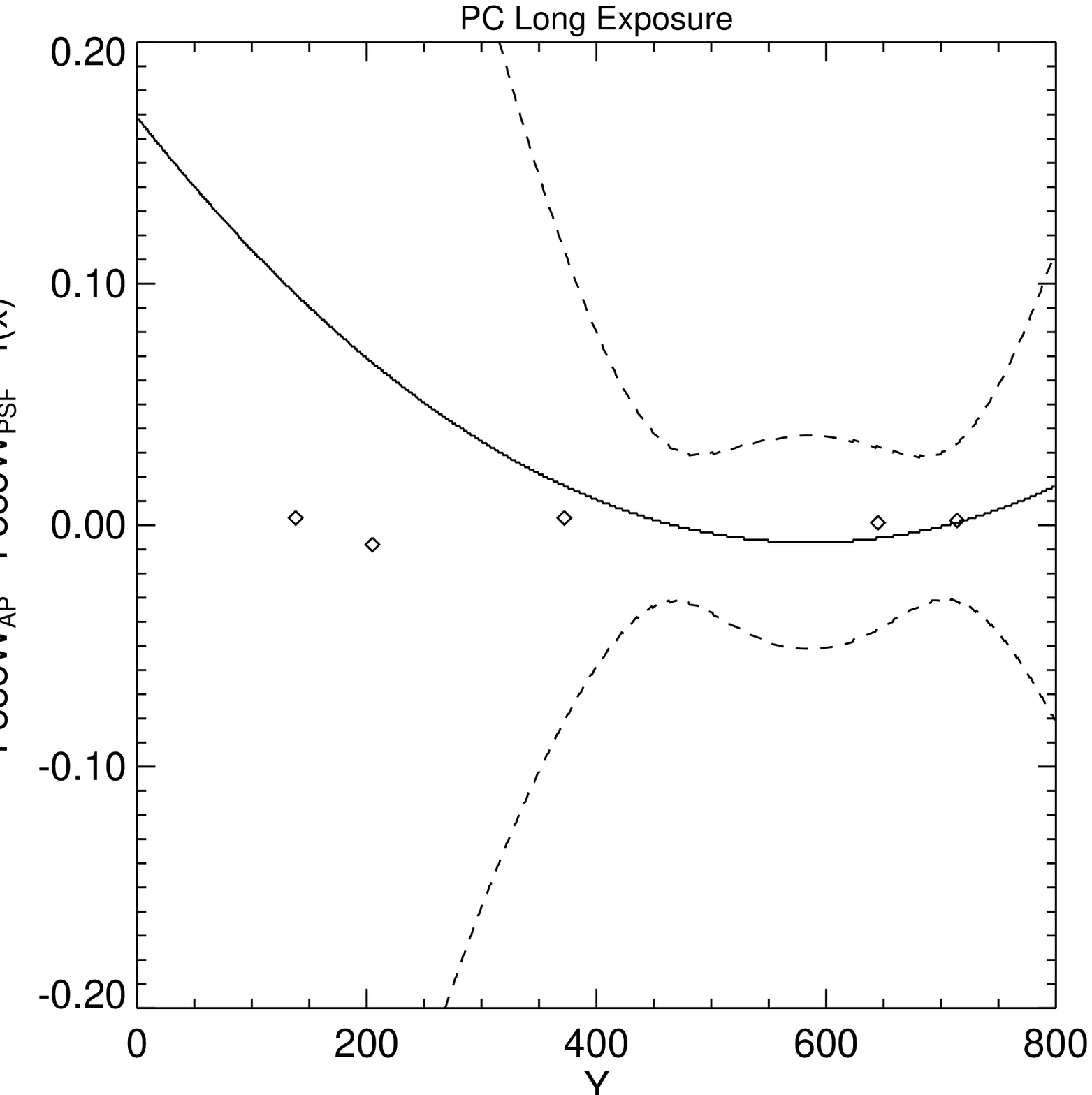}
\caption{An example showing the poor quality of the aperture corrections derived from the PC frames.  The solid lines are 2nd order polynomial fits while the dashed lines represent 1$\sigma$ deviations from the fit.}
\end{figure*}
\begin{figure*}
\plotone{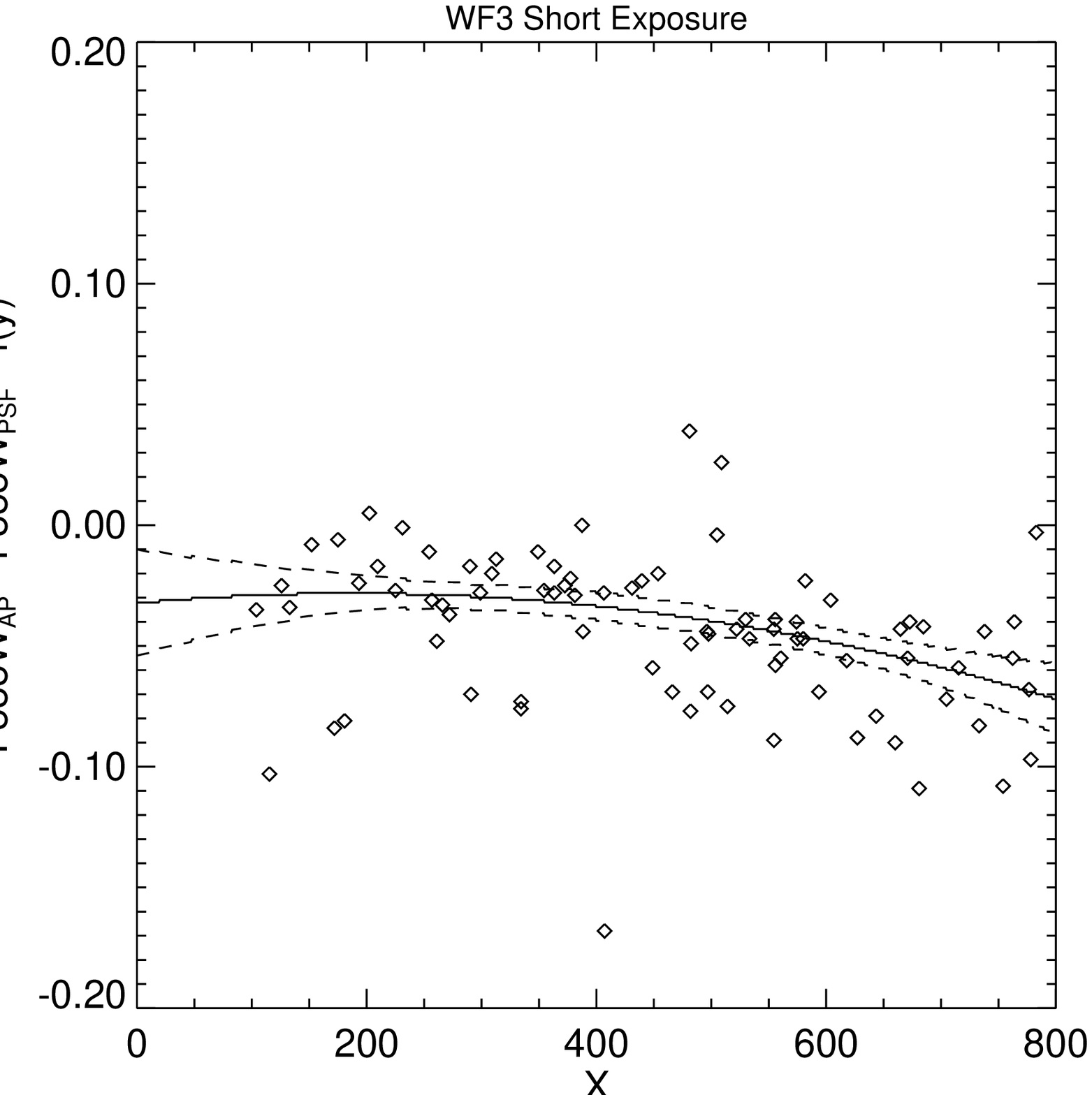}
\plotone{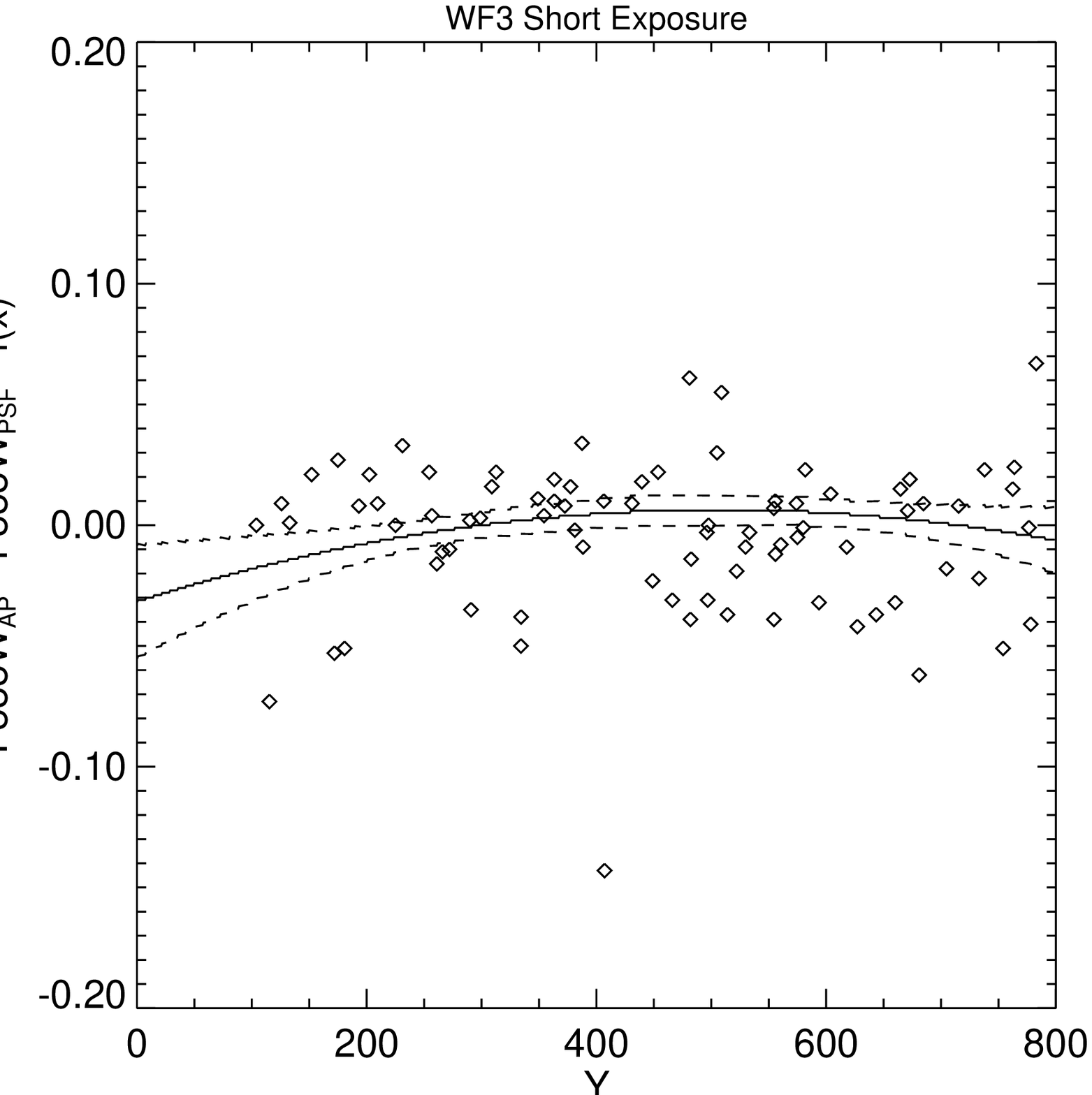}
\\
\plotone{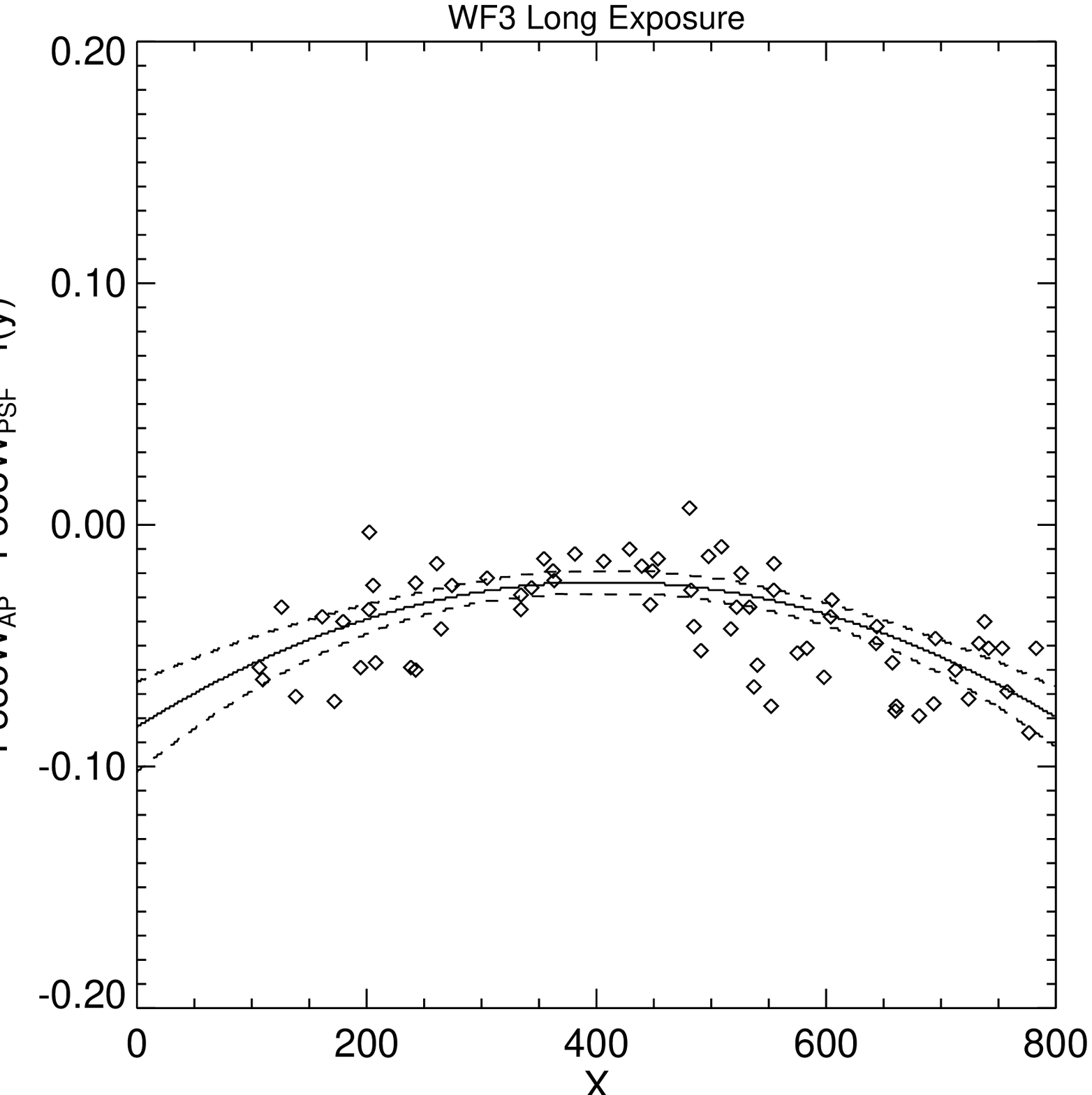}
\plotone{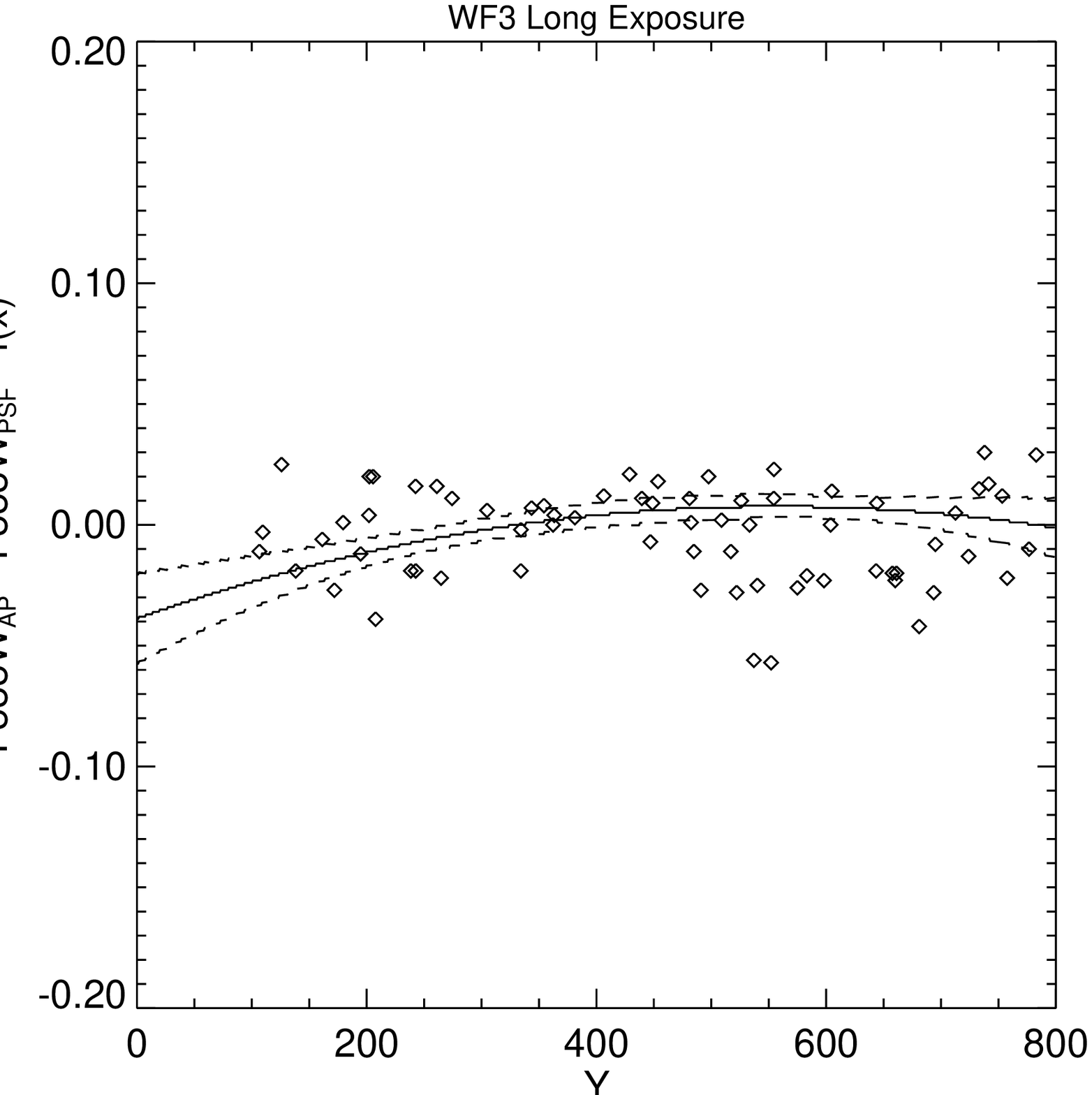}
\caption{An example showing the much higher quality of the aperture corrections of the less crowded WF frames.  The solid lines are again 2nd order polynomial fits while the dashed lines are 1$\sigma$ deviations.}
\end{figure*}
Finding the optimal size of the box within which the PSF is fit is a trade-off between the need to include pixels far from the core of the star for a proper fit to the background and the desire to exclude neighboring stars from the fit box.  The issue of crowding on the accuracy of the photometry is most critical in the crowded PC frames.  Fig. 2 shows the distribution of sky levels with position on the PC chip for two fit box sizes, 9 and 25 pixels, and the resulting CMDs of NGC 1754.  While the use of the larger 25-pixel fit box produces a much smoother sky distribution, the accompanying CMD shows much greater scatter.  We settled on the 9-pixel fit box for both the PC and WF frames, finding that this size produced adequate determinations of the background while avoiding, as much as possible, the effects of crowding.

When DoPHOT encounters objects that are significantly broader than the mean stellar profile, it makes a decision on whether to interpret the object as a single broad object (``galaxy") or as overlapping stars, in which case the object is split into multiple objects.  The decision is weighted by the parameter {\it stargalknob}, which we set to a high value because these crowded frames likely contain many more overlapping stars than background galaxies.  
We encountered some problems with DoPHOT classifying some bright yet unsaturated stars in the long exposure frames as broad, and splitting them into pairs.  Using the short exposure photometry for these cases solved this problem.

\begin{table}
\caption{DoPHOT PSF Parameters}
\begin{tabular}{lcccccc}
 & 
\multicolumn{3}{c}{F555W} & 
\multicolumn{3}{c}{F814W} \\
Chip &
$\beta_4$ &
$\beta_6$ &
FWHM &
$\beta_4$ &
$\beta_6$ &
FWHM \\
\hline
PC & 4.5 & 4.5 & 2.0 & 4.5 & 4.5 & 2.0 \\
WF2-4 & 2.2 & 2.2 & 1.2 & 2.5 & 2.5 & 2.0 \\
\hline
\end{tabular}
\end{table}

\subsection{Aperture corrections}
Because the model PSF is imperfect, the PSF-fitting photometry produces systematically different results from aperture photometry.  This systematic difference varies between different chips and filters.  Moreover, because the WFPC2 PSF varies with position on the chip, while DoPHOT assumes that the PSF is uniform with position, we applied an aperture correction that depends on position in the frame as well as the chip and filter used.  

For these aperture corrections, it was necessary to find several bright,
isolated stars in order to be able to do aperture photometry to the 0\farcs5
radius on which the Holtzman et al. (1995b) calibrations are based.  While we found a
few such isolated stars in the WF frames, there were
practically none in the PC frames.  We settled for using a PSF to subtract all
stars but a few bright ones and performing aperture photometry on the
subtracted image.
Because DoPHOT's PSF is uniform in position and leaves halos in the subtracted image, we chose to use PSFs kindly provided by Peter Stetson and DAOPHOT/ALLSTAR to produce the subtracted images.  These PSFs, derived from multiple WFPC2 observations of $\omega$ Cen, are ``perfect" PSFs, in that they average over time-dependent effects such as focus variations.  We found that ALLSTAR photometry using the $\omega$ Cen PSFs had greater scatter than our DoPHOT photometry, although the PSF-subtracted images are cosmetically cleaner. 

Choosing 200 bright stars to be left untouched in each frame, we used the ALLSTAR photometry to subtract the remaining stars.  We inspected the profiles of each of the 200 stars in the subtracted frame, discarding stars that had significant unsubtracted companions or deviant pixels within a 0\farcs5 radius.
With DAOPHOT's PHOTOMETRY routine, we performed aperture photometry on each of the 200 stars out to 0\farcs5.  After matching the aperture photometry lists with the DoPHOT photometry list, we fit the aperture corrections as a function of x and y on the chip with an appropriate polynomial.  Figs. 3-4 show sample fits of the aperture correction surfaces in the PC and WF3 frames for both long and short exposures.  In all of the WF
frames, a 2nd order polynomial surface fits well.  The residuals of the points around the mean surfaces are generally Gaussian-distributed with dispersions similar to those expected from errors in the combined aperture and PSF photometry, implying that we were successful in removing the stars neighboring to those used in calculating the aperture corrections.
However, in the PC frames
we find no clear dependence of the aperture correction on position, which we interpret as the 
result of the severe crowding and the difficulty of doing reliable aperture photometry in the PC frames.  Instead of using the questionable PC aperture photometry, we tested using an aperture correction based on Stetson's PSFs.  The artificial star tests used to generate the aperture correction surfaces based on Stetson's PSFs are described in Section 4.  To test the suitability of using an artificially generated aperture correction surface in the PC frames, Fig. 5 shows a sample comparison between the
WF2 aperture correction surfaces generated using our own data with artificial surfaces.  In most cases the agreement is good, but for some of the frames there are differences between the surfaces of $\sim$0.05 mag.  However, as we are uncertain of the PC aperture corrections generated from our own data at a level of $\ga$0.05 mag, we chose to use the artificially generated surfaces to correct the PC photometry.

\begin{figure}
\plotone{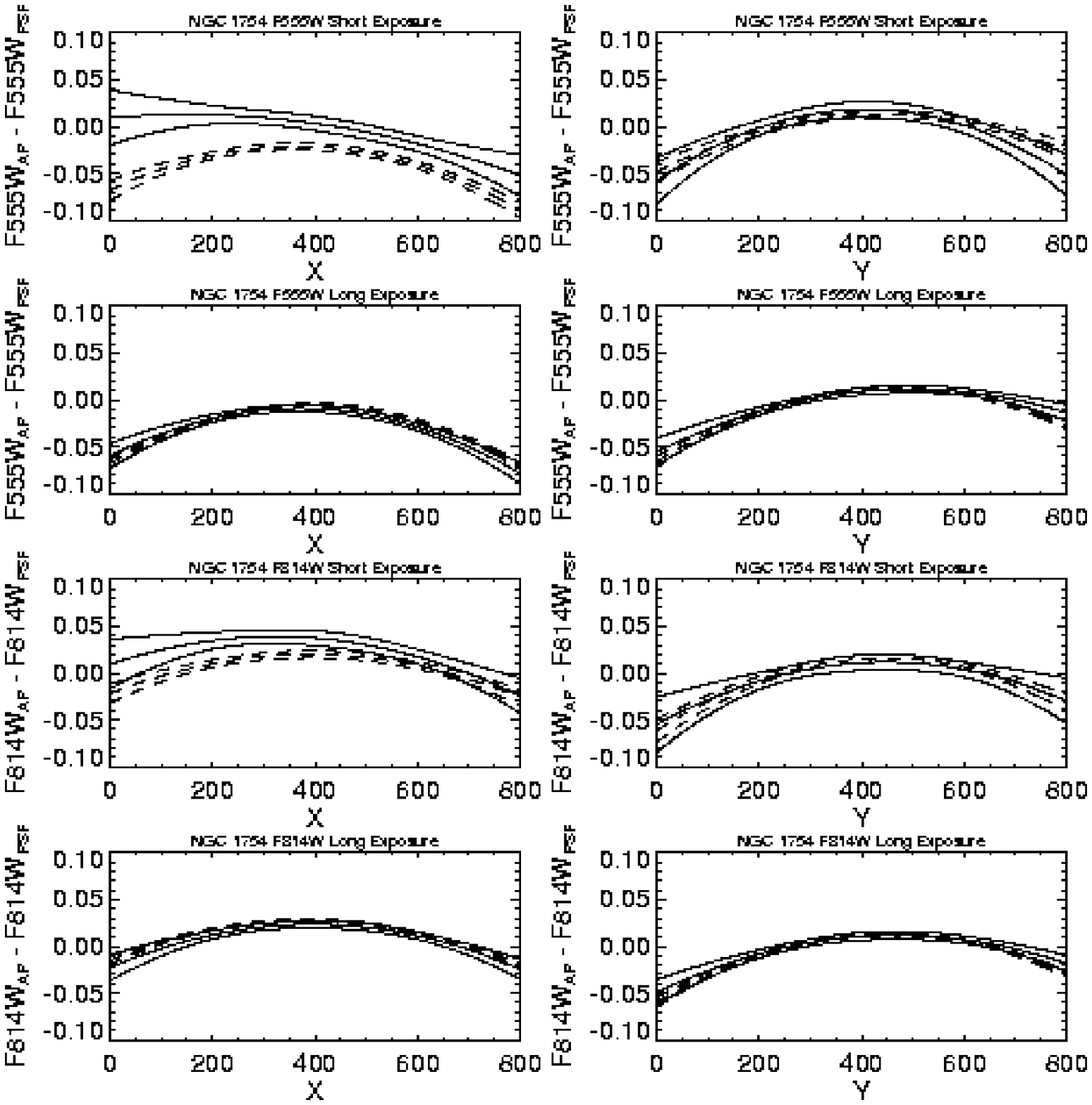}
\caption{An example of the agreement between aperture correction surfaces generated using Stetson's PSFs (dashed lines) with those derived from our WF2 frames (solid lines).  Both the 2nd order polynomial fits to the surfaces and the 1$\sigma$ deviations are shown.  The complete set of comparisons is available in Olsen (1998).}
\end{figure}

\begin{figure*}
\plotone{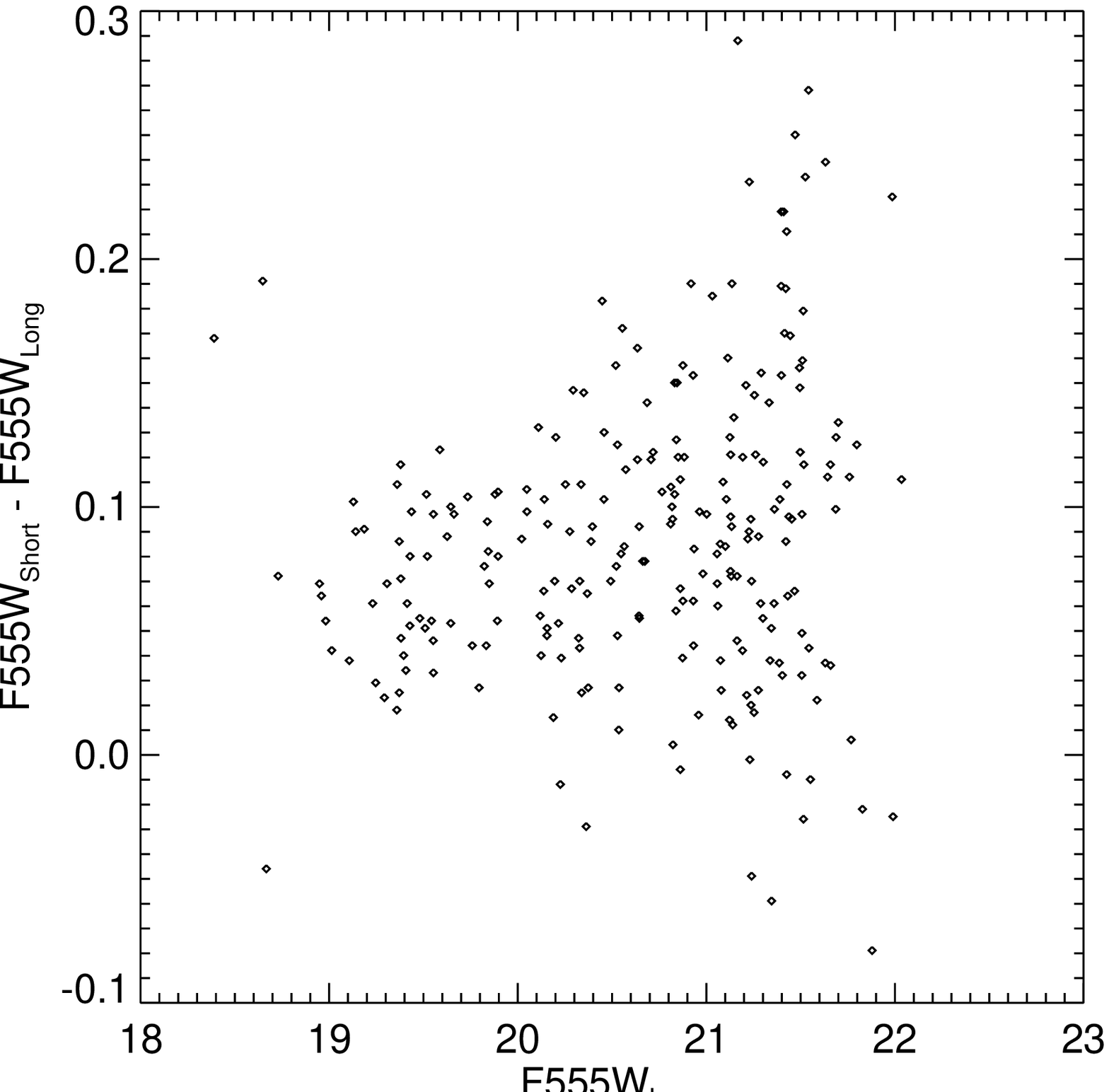}
\plotone{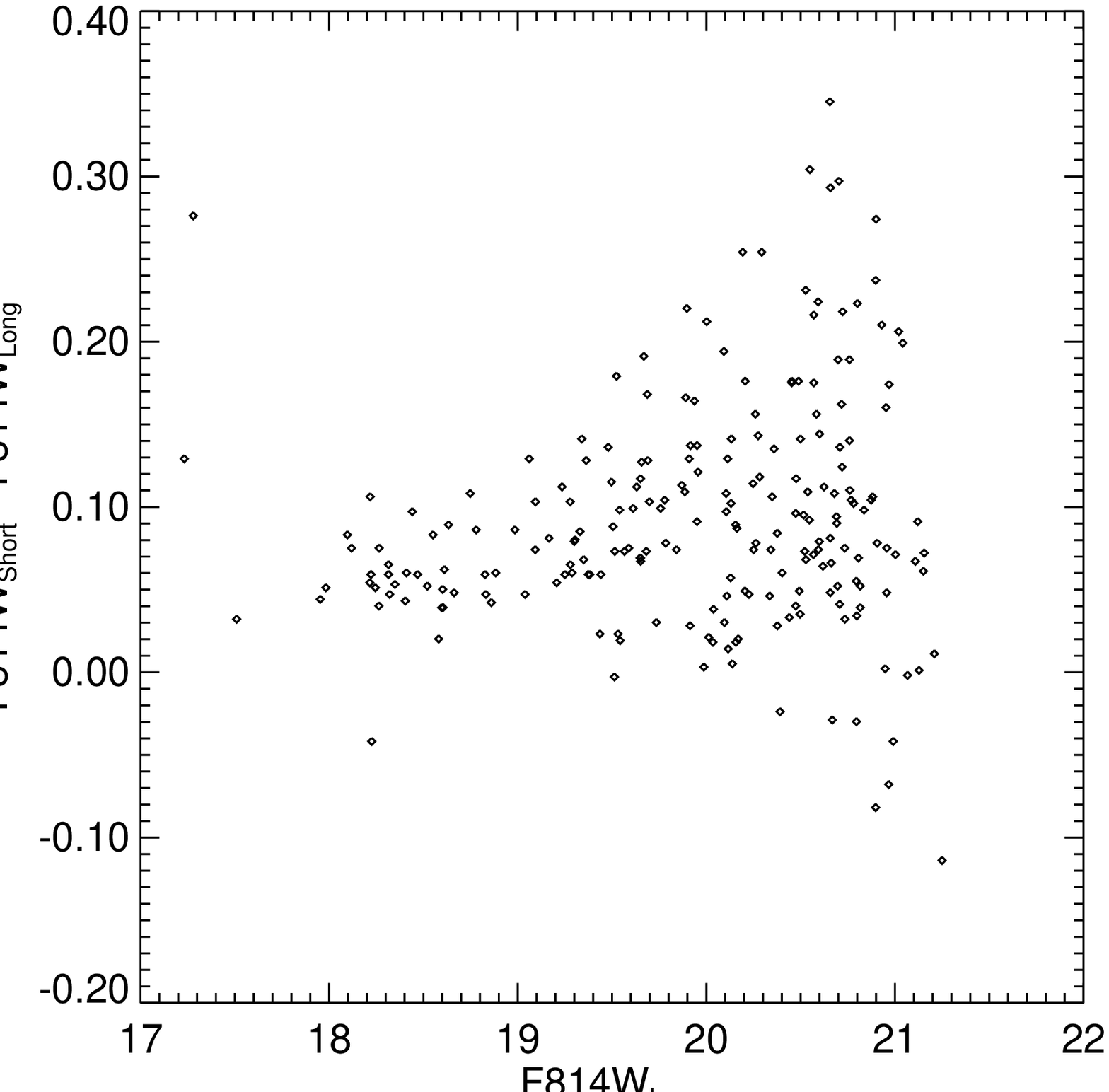}
\caption{A sample comparison of stellar magnitudes measured in the long (500, 600 s) WFPC2 exposures with the short (20 s) exposures, showing the CTE effect discussed by Whitmore \& Heyer (1997).}
\end{figure*}

\subsection{Offset between short and long exposure photometry}
In agreement with the report of Whitmore \& Heyer (1997), among others, on CTE
effects in WFPC2 photometry, we find an offset between magnitudes measured in
the long exposure frames compared to those measured in the short exposure
frames.  Fig. 6 shows a sample comparison.  The difference increases as the
magnitude of the star increases, reaching a maximum of $\sim$0.1 magnitudes at F555W = 21.5.  As
the offsets are due to a CTE effect which worsens at low background and count
levels, we chose to trust the long exposure photometry and apply a correction
to the short exposure photometry.  Within broad magnitude bins, we calculated
the average difference between the long and short exposure magnitudes and subtracted this difference from all of the short exposure magnitudes in the bin.  The use of broad
magnitude bins assumes that the
offset changes slowly with magnitude, with no erratic behavior hidden by the scatter in the plots.  We checked this assumption by comparing
the distribution of points around the average offset within each bin to the equivalent distribution generated by the artificial star tests described in Section 4.  The application of a Kolmogorov-Smirnov statistic shows that in all cases the distributions are nearly identical, implying that the assumption that the offset changes slowly with magnitude is well-justified.

\subsection{Final calibrated photometry}
We merged the short and long exposure photometry lists as
follows.  First, we removed stars near saturated and bad pixels from the
lists.  Stars within five pixels of a saturated pixel were assumed to lie in the
wings of the saturated star and were removed, while stars within 1.5 pixels of
a bad pixel were also removed.  Next, we matched the lists according to position of the stars.  When we plot the distribution of pixel offsets between stars found in both the short
and long exposure lists, we find that most of the distribution lies within 0.6
pixels of the origin, so matches outside this range were considered spurious.
For all matching pairs, we kept the photometry for the star having the smallest photometric error.
Sometimes, a star identified as single on the short exposure image was split into two stars by the long exposure reduction.  In these cases, we compared the photometric error of the single star with the error of the split pair member with the position closest to the single star, and kept the data for the star having the smallest error.  In most cases, this procedure rejected the photometry of the split pair member in favor of the single star and generally eliminated the previously mentioned difficulty of having bright stars split into two in the long exposures.
We also added stars found in the short exposure frames but
not in the long exposures to the final combined list, as these stars were generally saturated in the long exposures.

\begin{figure*}
\epsscale{1.5}
\plotone{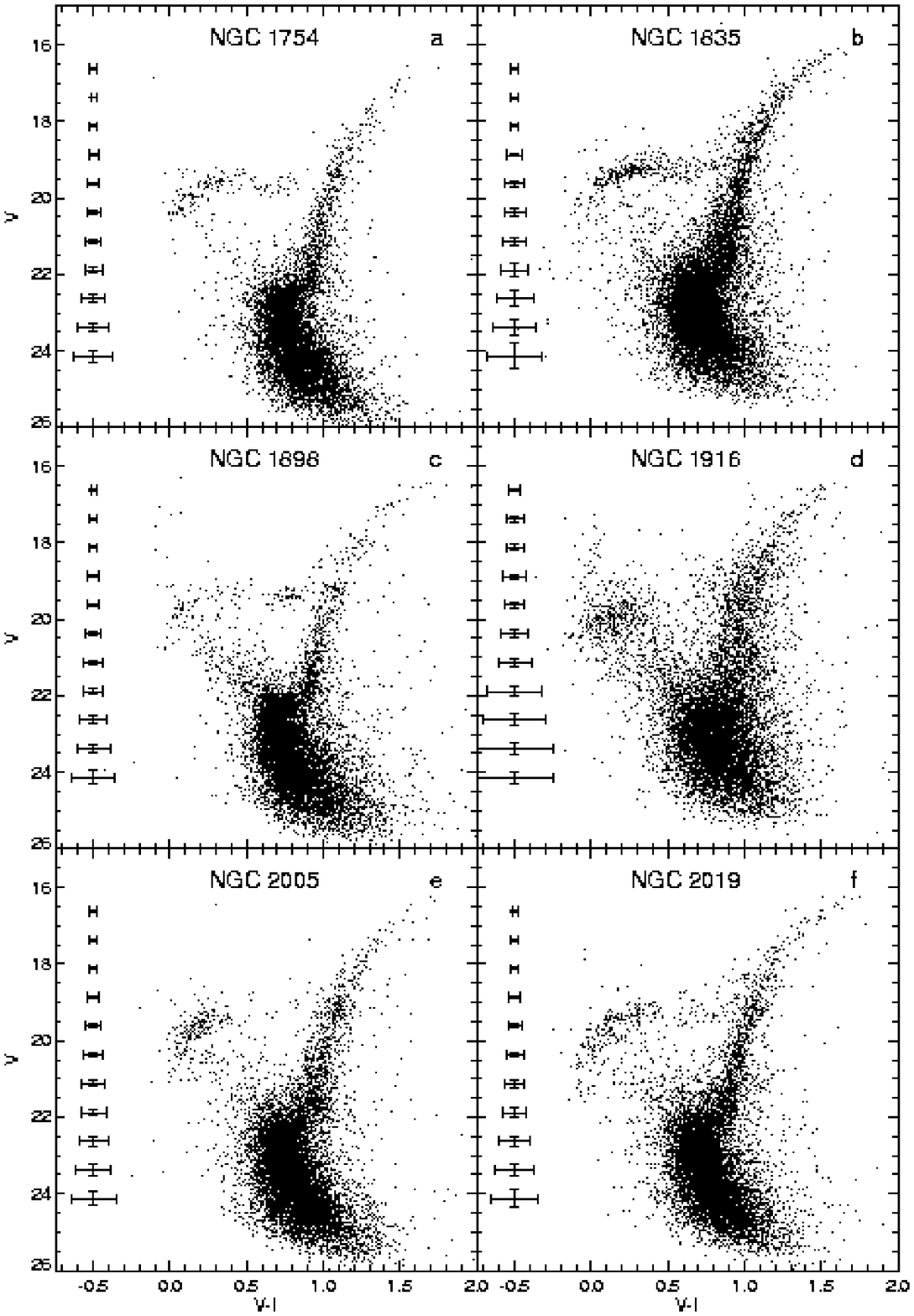}
\vspace{0.5cm}
\caption{Color-magnitude diagrams of the six old LMC globular clusters studied here.  No field star subtraction or other selective removal of stars has been performed and the diagrams are uncorrected for reddening.  The error bars are derived from the artificial star tests discussed in the text.}
\end{figure*}

Next, we matched the merged photometry lists from the two different filters, again using a 0.6 pixel matching
radius.  We kept only stars having a DoPHOT object type of 1 (``perfect'' star), 2 (``galaxy''), 3 (member of overlapping stellar group), or 7 (faint but measurable object) for the final list.  Using equation 9 and the zero points from table 10 of Holtzman et al. (1995b), we transformed our magnitudes to Johnson $V$/Kron-Cousins $I$, as this is the photometric system on which stellar evolutionary models are generally tabulated.  Figs. 7a-f show the CMDs generated from the PC photometry.  These CMDs are dominated by the cluster stars, but show some contamination due to field stars.  Table 4 contains a sample of our photometric data, the rest of which is available electronically through the Astronomical Data Center ({\tt http://adc.gsfc.nasa.gov/adc.html}).

\begin{table*}
\caption{LMC Globular Cluster Photometry}
\begin{tabular}{rccccccccc}
Star \# & 
X ($V$) &
Y &
$V$ &
$\sigma_V^a$&
$I$ &
$\sigma_I^a$ &
$V$ type$^b$ &
$I$ type$^b$ &
Removed$^c$ \\
\hline
NGC 1754 \\
1 & 386.11 & 100.57 & 19.3881 & 0.0380 & 18.3591 & 0.0520 & 10 & 10 & y \\
2 & 477.04 & 104.77 & 18.6639 & 0.0420 & 17.5141 & 0.0490 & 3 & 10 & \\
3 & 491.68 & 112.58 & 19.6949 & 0.0440 & 18.6781 & 0.0490 & 1 & 10 & \\
4 & 597.60 & 127.65 & 19.4229 & 0.0500 & 18.3281 & 0.0510 & 1 & 10 &   \\
5 & 233.08 & 129.43 & 19.6629 & 0.0460 & 19.3249 & 0.0540 & 1 & 1 &   \\
6 & 142.73 & 142.05 & 19.6841 & 0.0390 & 19.5021 & 0.0560 & 10 & 10 &   \\
7 & 753.78 & 153.25 & 19.3071 & 0.0550 & 18.1461 & 0.0500 & 30 & 10 &   \\
8 & 286.41 & 184.42 & 19.2699 & 0.0460 & 18.8031 & 0.0450 & 1 & 10 &   \\
9 & 105.07 & 203.56 & 19.6689 & 0.0540 & 19.2081 & 0.0520 & 3 & 10 &   \\
10 & 644.67 & 212.54 & 19.5009 & 0.0470 & 18.4271 & 0.0480 & 1 & 10 &   \\
\dotfill & & & & & & & & & \\
\hline
\end{tabular}
\begin{tabular}{l}
$^a$Photometric errors reported by DoPHOT \\
$^b$Where the short exposure photometry was used, the DoPHOT object type has been multiplied by 10.  \\
See text for explanation of object types. \\
$^c$Stars removed by field star cleaning procedure are marked with ``y". \\
\end{tabular}
\end{table*}

\subsection{Comparison with ground-based photometry}
In order to check the Holtzman et al. (1995b) zero points, we checked our calibrated WFPC2 photometry against our $V$ and $I$ CTIO 1.5-m observations of the cluster fields.  The
seeing in the 1.5-m images is generally $\sim$1\arcsec, corresponding to 22 pixels on the PC and 10 pixels on the WF chips.  Using approximate WFPC2 x and y positions of the ground-based stars, we identified stars from the 1.5-m and WFPC2 lists with positions matching within a radius equal to the radius of the 1.5-m seeing circle. 
For almost all of the stars appearing isolated on the 1.5-m images,
we found several stars within the seeing circle on the WFPC2 frames.  The
correct match was assumed to be with the brightest of the candidates in the WFPC2 frame.
We inspected each matching star by eye on the WFPC2 frame, and eliminated any star with
significantly bright companions within a 0\farcs7 radius from further consideration.  The remaining matches were used to establish the
comparison between the ground-based and WFPC2 photometry.

Because the four WFPC2 chips are essentially treated as different instruments
in our reduction, we expect that the average differences between the ground-based and HST magnitudes, measured on each chip and through both filters, will be randomly scattered, independent of chip or filter.  Instead, we find that they are strongly correlated.  We attribute this correlation to the extreme difficulty of
doing photometry in the severely crowded ground-based frames.  We also find that the difference between the ground-based and HST magnitudes is correlated with distance from the cluster center, which we believe is due to the increased crowding towards the cluster core.  After
removing the cluster-to-cluster variations, the ground-based and WFPC2 zero points are found to agree to within $\sim$0.1 mag.  We have thus not felt it necessary to adjust the Holtzman (1995b) zero points.

\begin{figure*}
\epsscale{1.}
\plotone{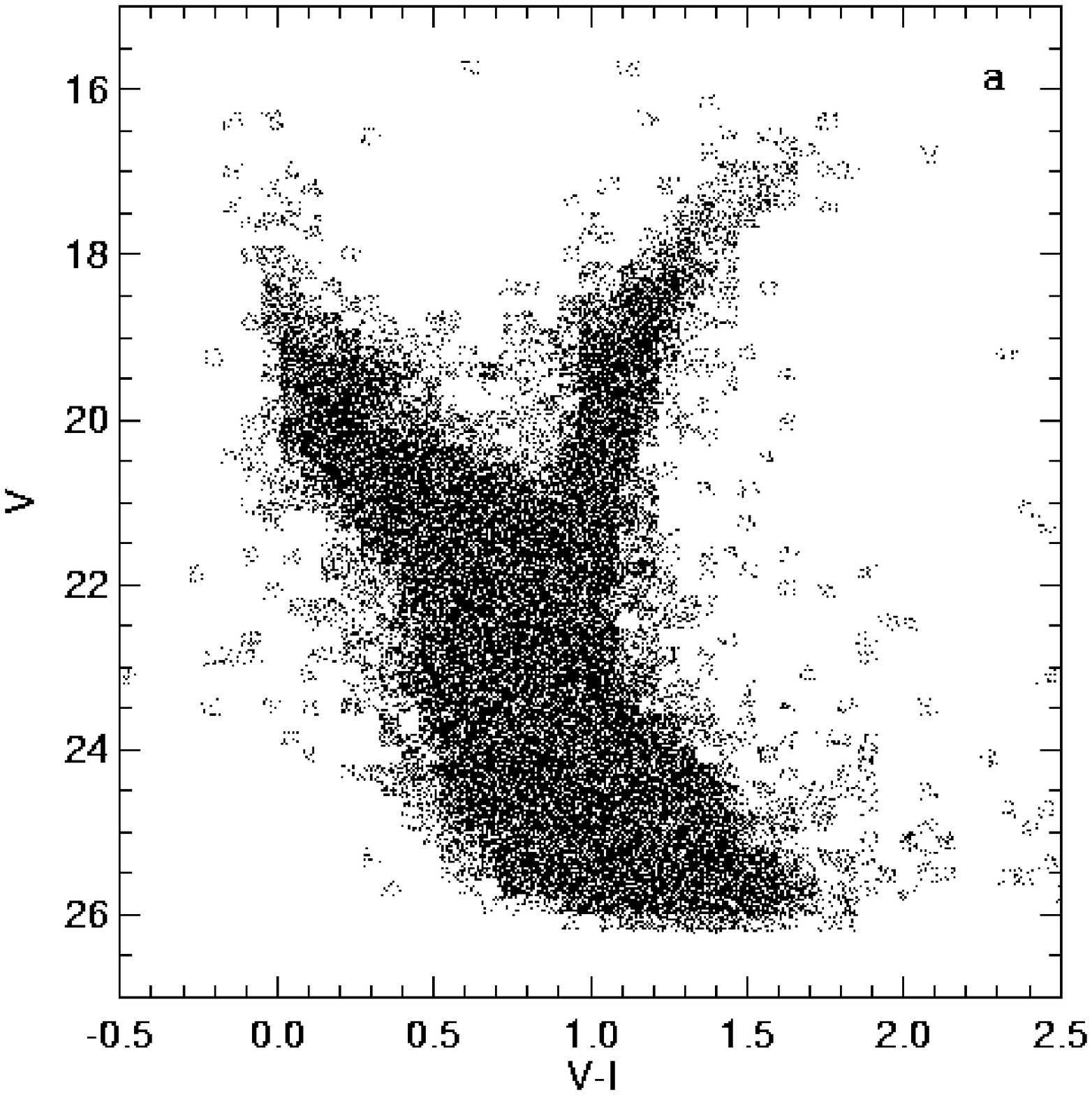}
\plotone{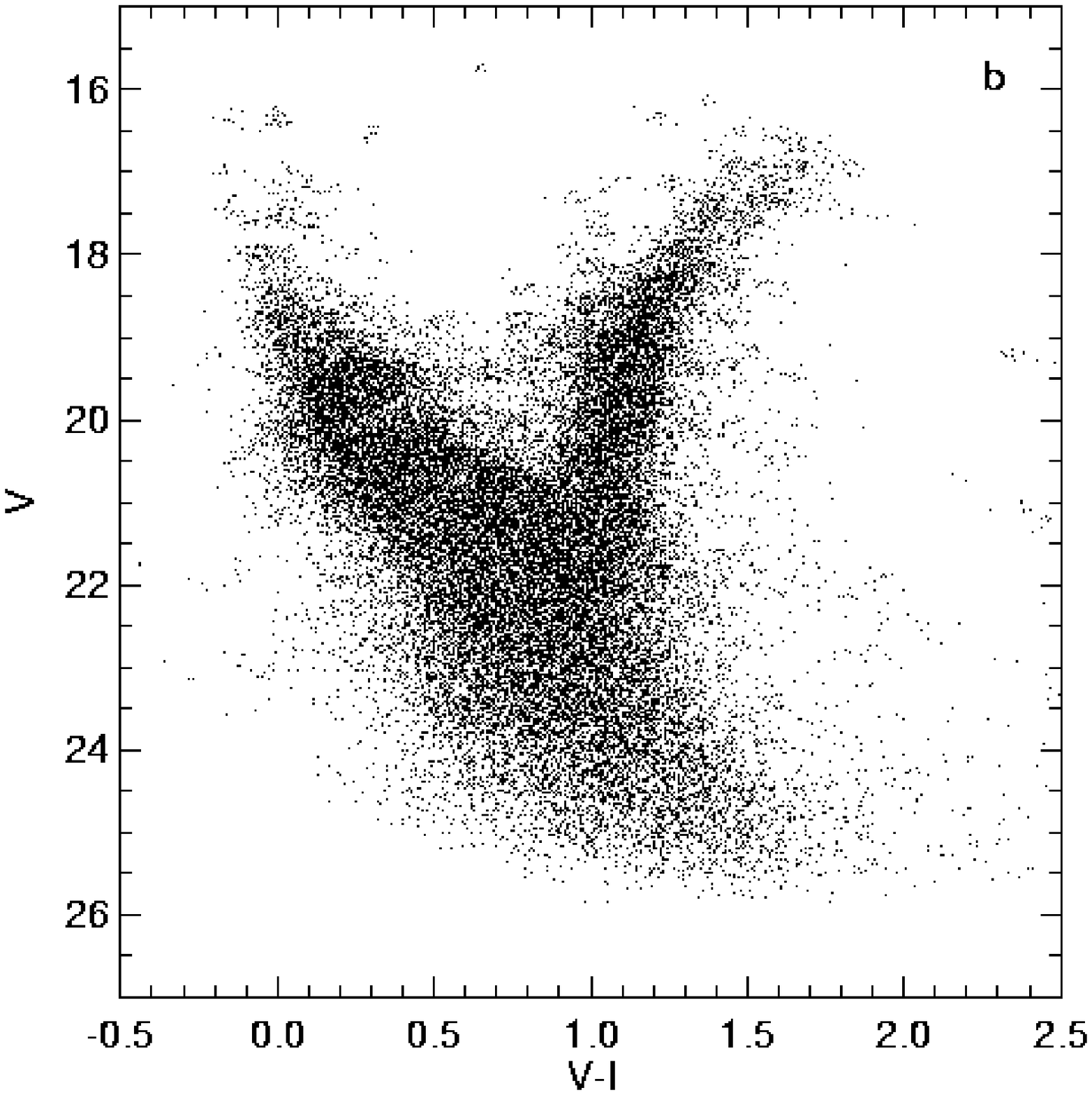}
\caption{An example of the input artificial star distribution ($a$) used in the tests discussed in the text and the colour-magnitude diagram recovered after addition to and processing through the WFPC2 images ($b$).}
\end{figure*}

\section{Artificial star tests}
\subsection{Generating the artificial star lists}
We conducted artificial star tests to study the completeness of the photometry, to examine the effects of crowding, and to aid in separating the field stars from the cluster stars.  The crowding and completeness effects are expected to be a function of magnitude, colour, and position in the frame.
Therefore, we used a set of artificial stars with magnitudes, colours, and positions distributed similarly to those of the stars in our CMDs.  We generated the magnitudes and colours of the input artificial
star sets by dividing our combined WF and PC CMDs into bins of 0.03-0.06 magnitudes in $V-I$ and
0.1-0.2 magnitudes in $V$.  In each bin, we chose a number of random points to represent the artificial stars.  The number of artificial stars in a bin was chosen to be 10 times the number of observed stars in that bin, with the constraint that no bin should contain more than 100 artificial stars.  Fig. 8a
shows a sample input artificial star CMD.  We distributed the positions of the artificial stars randomly, weighting the distribution by the stellar density profile.  We determined the stellar density profile 
by fitting the completeness-corrected data to a King model plus a constant background, with completeness corrections measured from an earlier artificial star experiment in which the artificial stars were uniformly distributed in position.  Although useful for estimating the completeness, this initial experiment was not helpful in statistically subtracting the field and cluster CMDs from one another because of insufficient coverage in $V-I$.  We refit the King models later from the updated completeness corrections of the new artificial star experiments. 

Using the previously
mentioned PSFs obtained from Stetson, we distributed the artificial star sets
over multiple copies of the original images, with the density of artificial stars in a given image amounting to $\sim$5 per cent of the original stellar density.
Each artificial star was placed in both the long and short exposure frames and in both filters.
In all, $\sim$6000 artificial star images were generated.
The frames containing the artificial stars were reduced
in exactly the same manner as the original images, although it was, of course, unnecessary
to perform any CTE correction to the short exposure photometry.  To recover
the detected artificial stars, we searched the DoPHOT output lists for stars with positions matching within 0.6 pixels to those of stars in the input artificial star list.  The list of recovered stars was similarly compared with the list of stars from the original image.  In cases where the recovered artificial star also matched a star in the original image, we considered the artificial star lost if the recovered magnitude was closer to the magnitude of the real star than to the magnitude of the artificial star.  As with the real photometry, artificial stars placed too closely to bad or saturated pixels were considered lost.  We merged the lists of recovered artificial stars placed in the short and long exposure images, and matched the merged lists from the two filters.  Fig. 8b shows the CMD of the stars recovered from the distribution in Fig. 8a.  

\subsection{The effects of crowding and incompleteness}
Crowding and incompleteness affect our analysis in a number of ways.  Because the differences in crowding between the cluster-dominated PC frames and the field star-dominated WF frames are large, we need to quantify the crowding effects in order to separate the field and cluster stars.  Proper field and cluster star separation also depends on our fit to the cluster profile.  To measure this profile from the stellar density distribution, we need to measure the completeness as a function of both position and magnitude.  Finally, because the cluster ages that we measure depend sensitively on the accuracy of the photometric scale, we need to know whether crowding introduces any changes in the colour or magnitude scales.  Based on our understanding of the crowding and completeness through the artificial star tests, we will be able to select those stars with the most reliable photometry for conducting our analysis of the CMDs.

Fig. 9 shows a sample completeness surface calculated as a function of position and $V$ magnitude.  The surface represents the probability that a star of a given position and actual $V$ magnitude would be found in our CMDs.  The surface shows how the completeness and associated limiting magnitude are strong functions of position near the cluster center.

\begin{figure}
\plotone{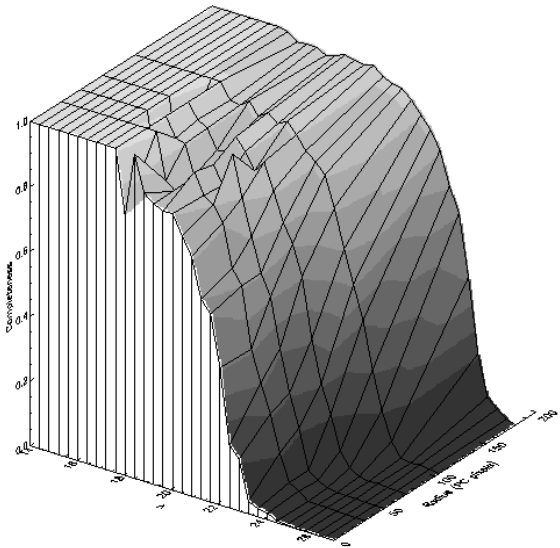}
\caption{A sample of the calculated completeness as a function of $V$ and radius from the cluster center.  The gray shadings are levels of constant completeness fraction.  At bright $V$ magnitudes, the completeness has been set to 1 where no artificial stars were placed to calculate the completeness.}
\end{figure}

We used the artificial star tests to evaluate both random errors and systematic colour and magnitude shifts in the C-M plane.  Fig. 10 shows an example of the random errors and systematic shifts in the photometry.  While the size of the random errors depend mostly on $V$ with very little, if any, colour dependence, the systematic shifts depend on both $V-I$ and $V$.  At faint $V$ magnitudes, this dependence of the shifts on both $V$ and $V-I$ can be explained by the existence of a magnitude limit in both the $V$ and $I$ filters.  At brighter magnitudes, however, there remains a tendency for blue stars to be recovered redder than their true colours.  This is likely an effect of crowding with the predominantly red cluster stars in the PC frame.
\begin{figure}
\epsscale{1.0}
\plotone{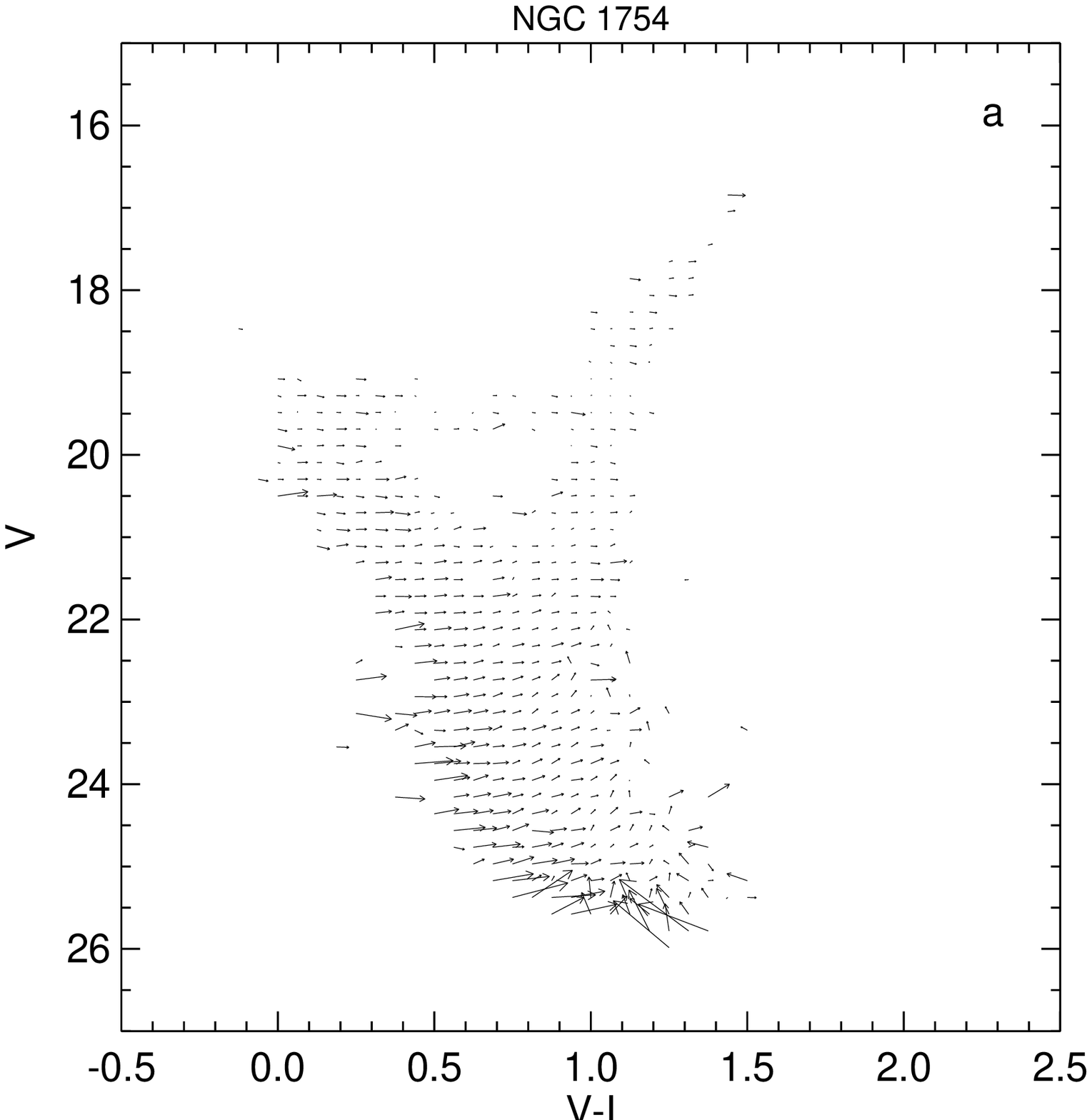}
\\
\plotone{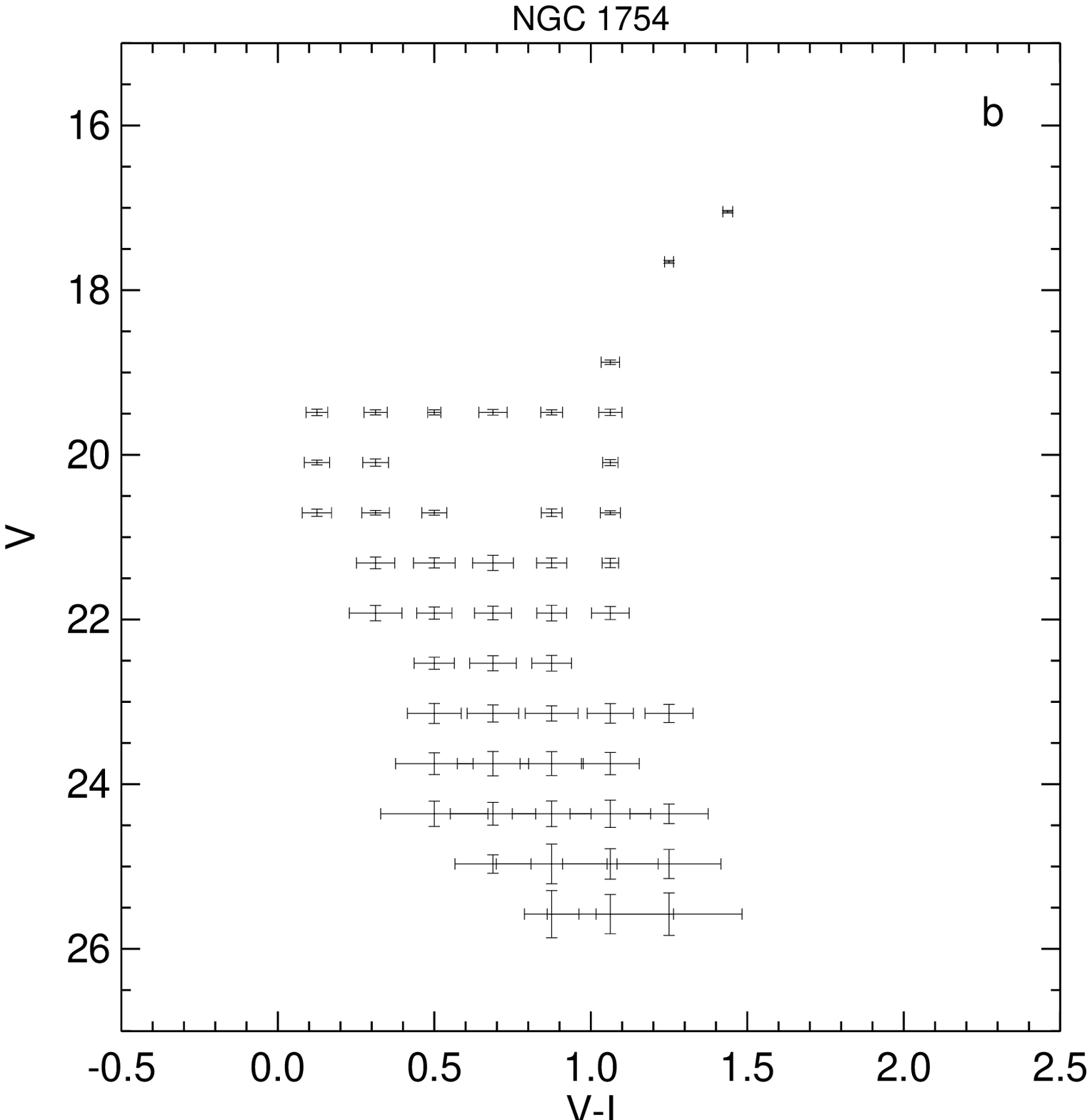}
\caption{Systematic and random photometric errors calculated through artificial star tests for the example of NGC 1754.  In ($a$), the arrows show the mean direction and distance over which stars originating in the bin at the tail of the arrow drifted in the colour-magnitude plane.  In ($b$), the error bars show the 1$\sigma$ robust standard deviations of the recovered colours and magnitudes of the artificial stars in each bin.  For similar plots for the other clusters, see Olsen (1998).}
\end{figure}

Our WFPC2 frames are a mix of cluster and background field stars, with no portion of the images containing only one type of star.  However, because the cluster and field stars are distributed differently, we can still statistically separate the two populations.  We are further aided by the fact that the PC frames contain more cluster stars than field stars, vice versa in the WF frames. Our approach is to use the WF frames to remove field stars from the PC frame, and use the cleaned PC frame to remove cluster stars from the WF frames.  After a few iterations of this procedure, we achieve satisfactorily clean cluster and field CMDs.

Our cleaning procedure accounts for the different crowding conditions and the different areas subtended by the cluster and field star distributions in the PC and WF frames.  The following equation describes the number of contaminating background stars expected in the frame containing the stars of interest:
\begin{equation}
N(V,V-I) = \frac{\Lambda^p(V,V-I)}{\Lambda^c(V,V-I)}\frac{\int F(\vec{r})dA^p}{\int F(\vec{r})dA^c}D^c(V,V-I)
\end{equation}
where $\Lambda^p(V,V-I)$ is the crowding surface of the primary frame, $\Lambda^c(V,V-I)$ is the crowding surface of the background frame, $F(\vec{r})$ is the spatial distribution of the background stars, $dA^p$ and $dA^c$ are area elements, and $D^c$ is the distribution of stars in the CMD in the background frame.  We assume the field stars are uniformly distributed, so that $F(\vec{r})=$constant and the integrals reduce to the areas of the frames.  For the cluster stars, we choose a King (1966) model to represent $F(\vec{r})$.  The next section discusses these King model fits.

\subsection{Cluster profiles}

We determined the centers of the clusters through an algorithm developed by King and described in Mateo (1987).
Briefly, the algorithm finds the center of symmetry of
the clusters by extracting subimages on a grid of points in the image containing the cluster
and correlating each subimage with itself, rotated 180 degrees.  The value of
the correlation is stored at each grid point, resulting in a surface whose
maximum represents the center of symmetry.  We measured the maximum by fitting
this surface to a second order polynomial and computing the
vertex.  The algorithm worked well for the rich clusters, but was somewhat
time-consuming and more difficult in the poorer clusters, particularly NGC 1898.

\begin{figure}
\epsscale{1.0}
\plotone{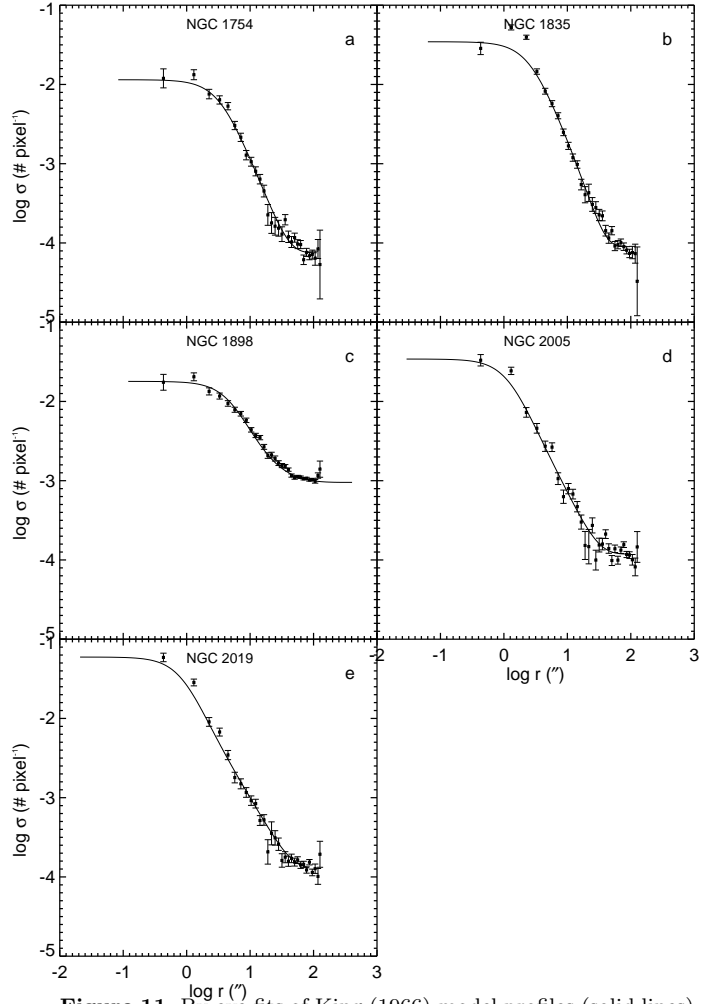}
\vspace{0.2cm}
\caption{By-eye fits of King (1966) model profiles (solid lines) to the completeness-corrected stellar surface density profiles of each cluster (open squares).  The error bars are due to counting statistics and uncertainty in the completeness corrections.}
\end{figure}
\section{Cleaning background stars from the CMDs}

With the cluster centers established, we calculated the radial stellar distributions of each cluster.
We transformed the coordinates of all 
the stars to a reference frame centered on the cluster and scaled to the size of PC
pixels.  We set up thirty logarithmically spaced annuli around the center of each
cluster out to the edges of the WF chips and calculated the area of the portion of each
annulus covered by the WFPC2 chips. We then summed up the number of
stars within each annulus, divided the sum by the completeness
appropriate to the stars' positions and magnitudes, and divided by the
appropriate area.  Because the magnitude limit is a function of position, we tried to restrict the radial distribution to contain only stars with completeness $\ga$ 50 per cent at all radii.  In order to have enough stars to define the profiles of NGC 1835, NGC 2005, and NGC 2019, however, we were forced to include stars with completenesses as low as 20 per cent.
 
We fit King (1966) models with variable amounts of background to the radial distributions to determine the relative contribution of the cluster stars in each chip.  First, we 
estimated the peak density from the profile close to
the center and set the core radius, $r_c$ to the radius at which the density reaches $\sim1/2$ the
peak radius.  We then tried different variations of $r_c$, concentration, and
background level and chose the best fit by eye.  We note that for NGC 1835, the innermost three points do not give a good fit, probably because of incompleteness problems near the cluster center.  Figs. 11a-e show the observed background-subtracted profiles for each cluster and the adopted fits.  The parameters of these fits are included in Table 6.

\begin{figure*}
\epsscale{1.45}
\plotone{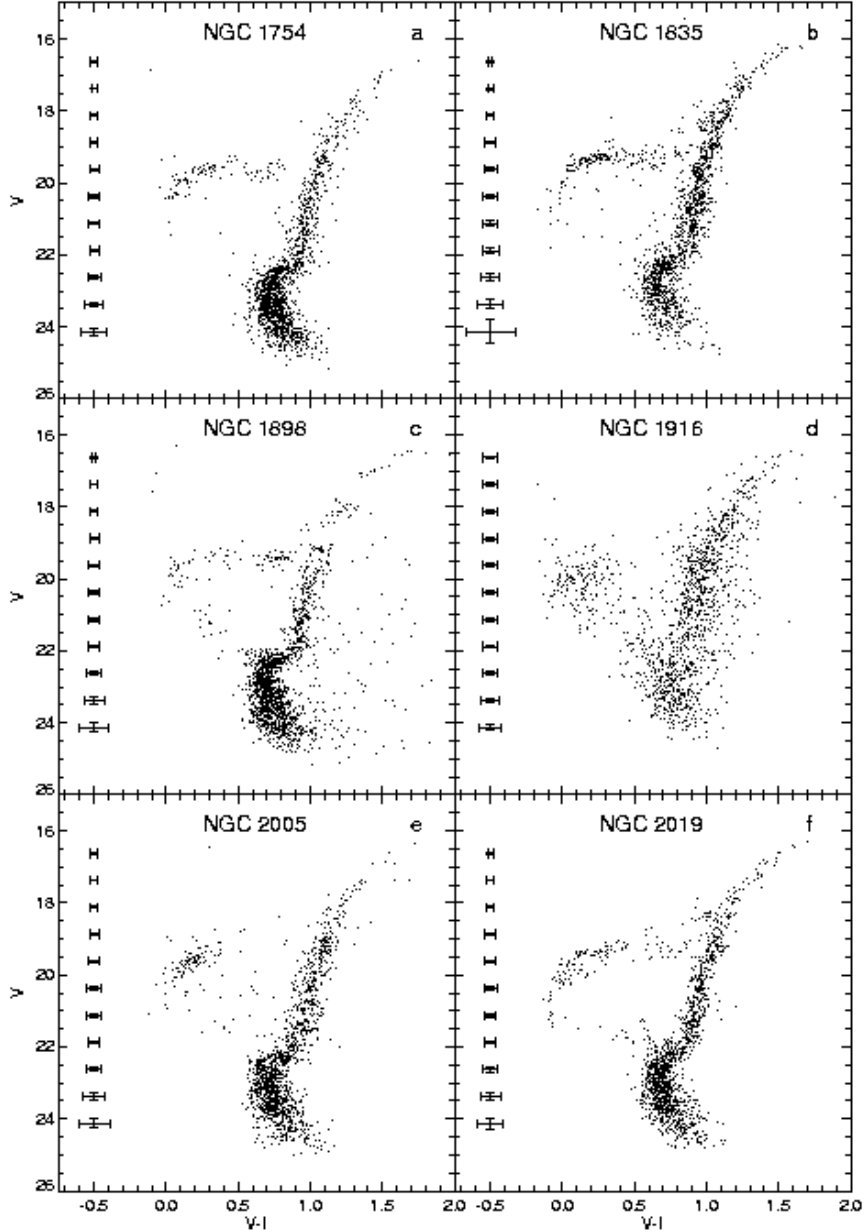}
\vspace{0.5cm}
\caption{Color-magnitude diagrams of the LMC clusters after selection of stars of DoPHOT object type 1 with $r \ga 9\arcsec$ 
from the cluster centers.  For all of the clusters except NGC 1916, field stars have been statistically removed.  The error bars are those calculated from the artificial star tests discussed in the text.}
\end{figure*}

\subsection{Removal of stars}
We began by removing a statistical sample of field stars chosen from the combined WF CMDs from the cluster-dominated PC CMD.  Equation 1 becomes:
\begin{displaymath}
N^{PC}(V,V-I) = \frac{\Lambda^{PC}(V,V-I)}{\Lambda^{WF}(V,V-I)}\frac{A^{PC}}{A^{WF}}D^{WF}(V,V-I)
\end{displaymath}
$N^{PC}(V,V-I)$ represents how the stars in the WF CMD would be distributed in the PC CMD, and are the stars we wished to subtract.  However, because some areas of the CMDs are sparsely populated, not all of the bins in the PC CMD for which 
$N^{PC}(V,V-I)$ is positive contain stars.  Therefore, rather than simply subtracting the $N^{PC}(V,V-I)$ stars from the appropriate bins, we subtracted stars based on the error ellipse at each point.  At each point $(V,V-I)$, we gave the surrounding stars weights based on the error distribution at $(V,V-I)$, and chose the stars to be subtracted randomly from this weighted distribution.  Stars outside the 3-$\sigma$ error ellipse were given zero weight and were not considered.  If no stars laid within the 3-$\sigma$ ellipse, we subtracted the nearest available neighbor.
We then used the cleaned PC CMDs to subtract cluster stars from the WF CMDs.  The relevant equation is:
\begin{displaymath}
N^{WF}(V,V-I) = \frac{\Lambda^{WF}(V,V-I)}{\Lambda^{PC}(V,V-I)}\frac{\int F_{\rm King}(r)dA^{WF}}{\int F_{\rm King}(r)dA^{PC}}
\end{displaymath}
\begin{displaymath}
\hspace{6cm}\times D^{PC}(V,V-I)
\end{displaymath}
where $F_{\rm King}(r)$ is the King model fit from the completeness-corrected stellar density distribution and $D^{PC}(V,V-I)$ is the cleaned PC CMD.  We used the same procedure described above to remove stars from the WF CMDs, producing CMDs with mostly cluster stars removed.

Because there are both cluster and field stars in both the PC and WF CMDs, we necessarily removed some cluster stars from the PC CMDs and some field stars from the WF CMDs in the first iteration of the subtraction procedure.  Repeated iterations improved the separation, producing statistically cleaner cluster and field star CMDs.  We found that the procedure converges in $\sim$5 iterations, as expected by the measured contamination levels and a simple calculation of the convergence rate.  

\section{Analysis of cluster CMDs}

\subsection{Overview}
Figs. 12a-e show the cleaned cluster CMDs.  Most field stars have been removed, although a few remain between the MSTO and the HB.  Overall, however, the cleaning process makes it much easier to see the location of the MSTO.  
The CMDs all clearly show that the clusters are old, with MSTOs at V$\sim$23 and blue HB stars along with red ones.  Comparing our CMD of NGC 1754 with the ground-based CMD of Jensen, Mould, \& Reid (1988), we see the immense difference that the resolution of HST makes in these crowded fields.  While Jensen et al.'s (1988) effort was valiant, their misidentification of the field main sequence and the HB of NGC 1754 with the cluster main sequence led them to suggest an age of only 0.8 Gyr for the cluster.

It is tempting to suggest that some of the blue stars just above the MSTO in NGC 1898 and NGC 2019 may be true blue stragglers.  However, considering the large numbers of young MS field stars in this region and the uncertainty in the statistical cleaning procedure, we make no such claim here.  With the exception of NGC 2005 and possibly NGC 1916 (Fig. 7d), all of the clusters have HB stars both to the blue and to the red of the instability strip.  Both the NGC 1835 and NGC 2019 CMDs show a few AGB stars, as do possibly the other clusters.

NGC 1898 is unique in containing a large number of stars to the red of the primary fiducial sequence.  As we were puzzled by these stars, we selected them on the CMD and examined their locations in the image.  Figs. 13a\&b show the selected stars and their locations on the image.  The majority of the stars lie in an area of the sky that contains relatively few stars--this plus the anomalously red colours of these stars indicates that the area is a patch of heavy obscuration.  Amazingly, the rest of the cluster suffers only a modest of amount of reddening, comparable to that of the other clusters.
 
\begin{figure}
\epsscale{1.0}
\plotone{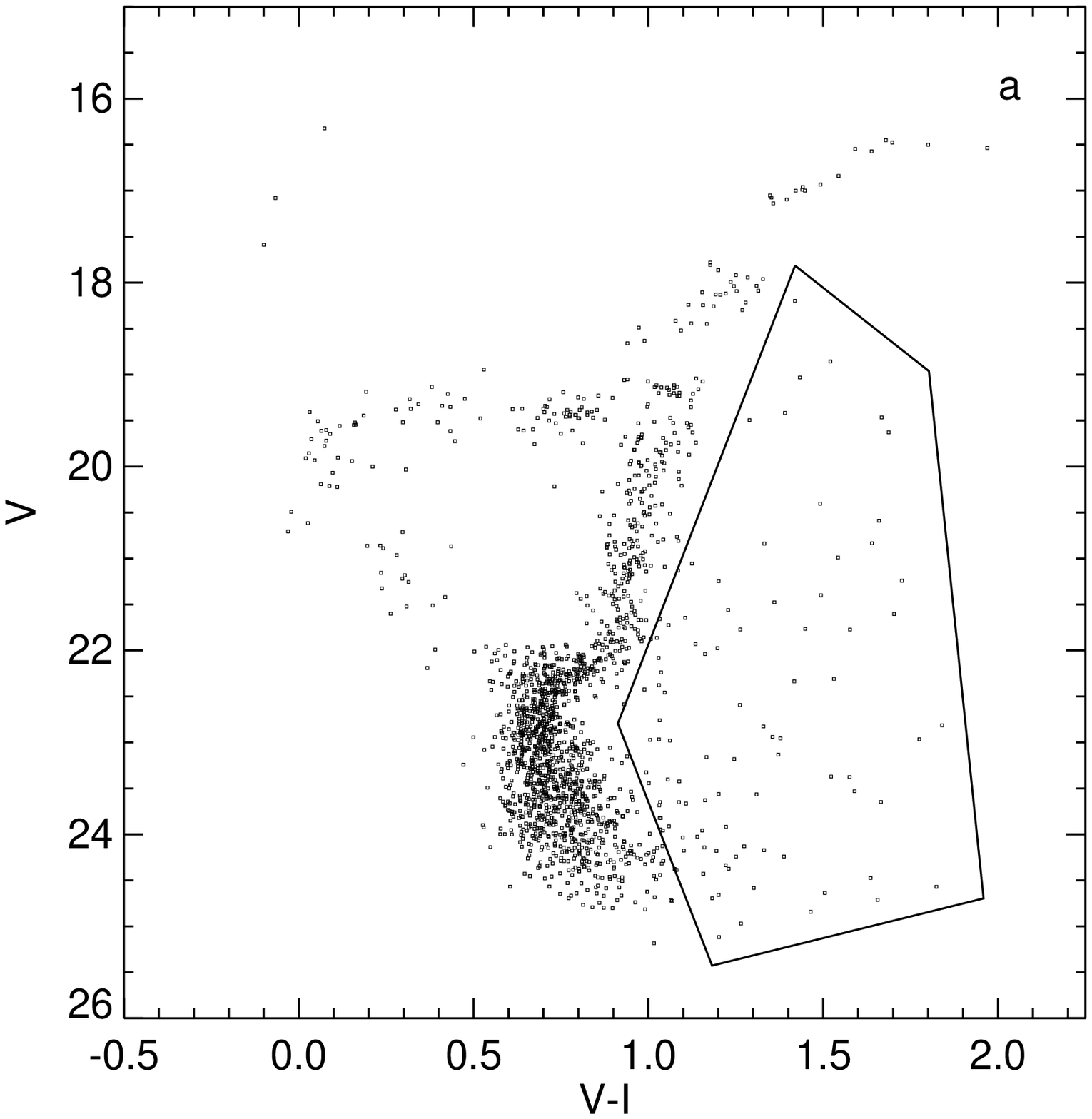}
\\
\plotone{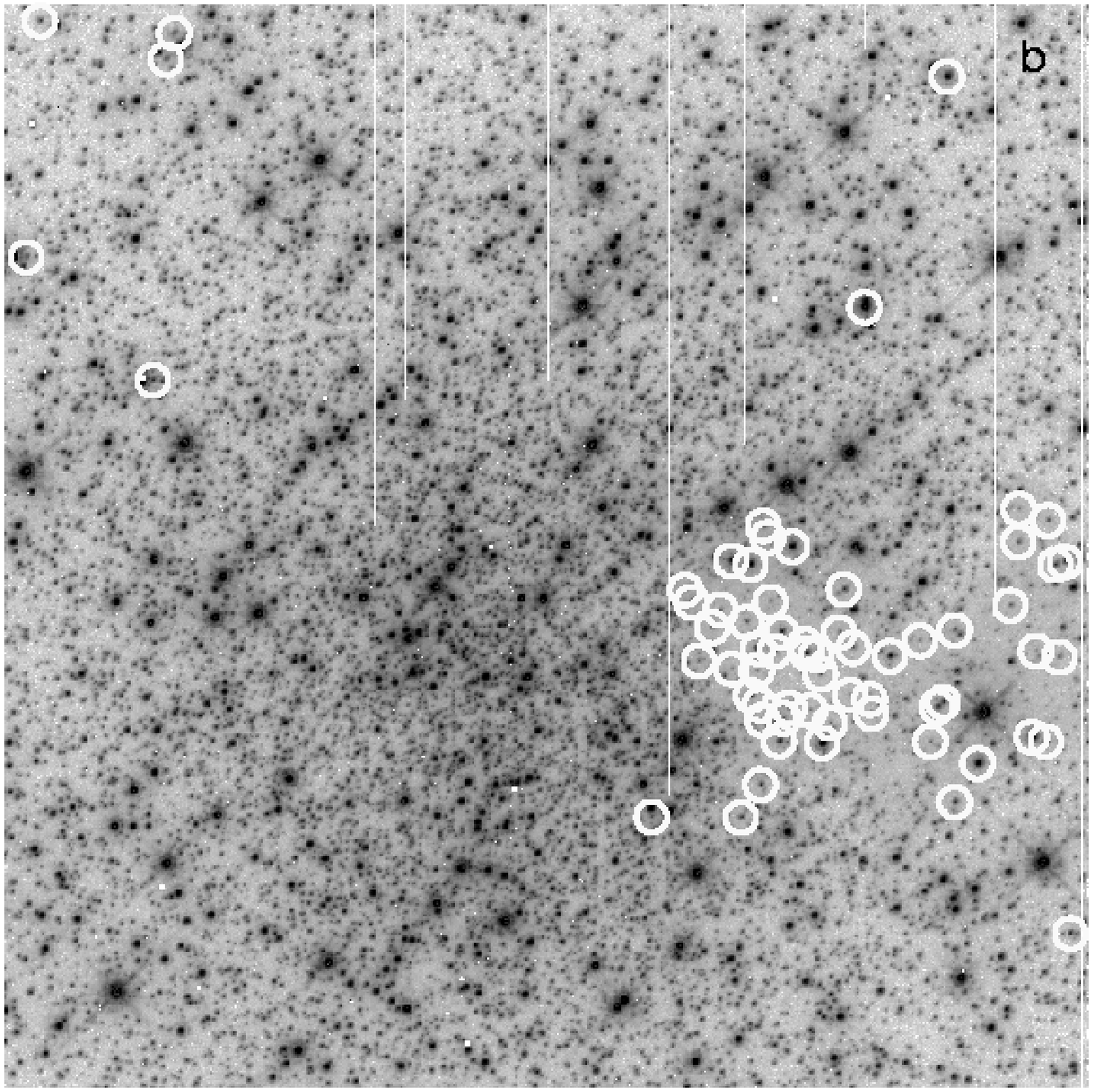}
\caption{($a)$ The peculiarly red stars in the colour-magnitude diagram of NGC 1898 were selected within the bounding polygon shown and their positions displayed on the image in ($b$).  The selected stars are heavily concentrated towards a region of the image with noticeably lower star density, and are therefore very likely highly reddened.}
\end{figure}

\begin{figure*}
\epsscale{1.7}
\plotone{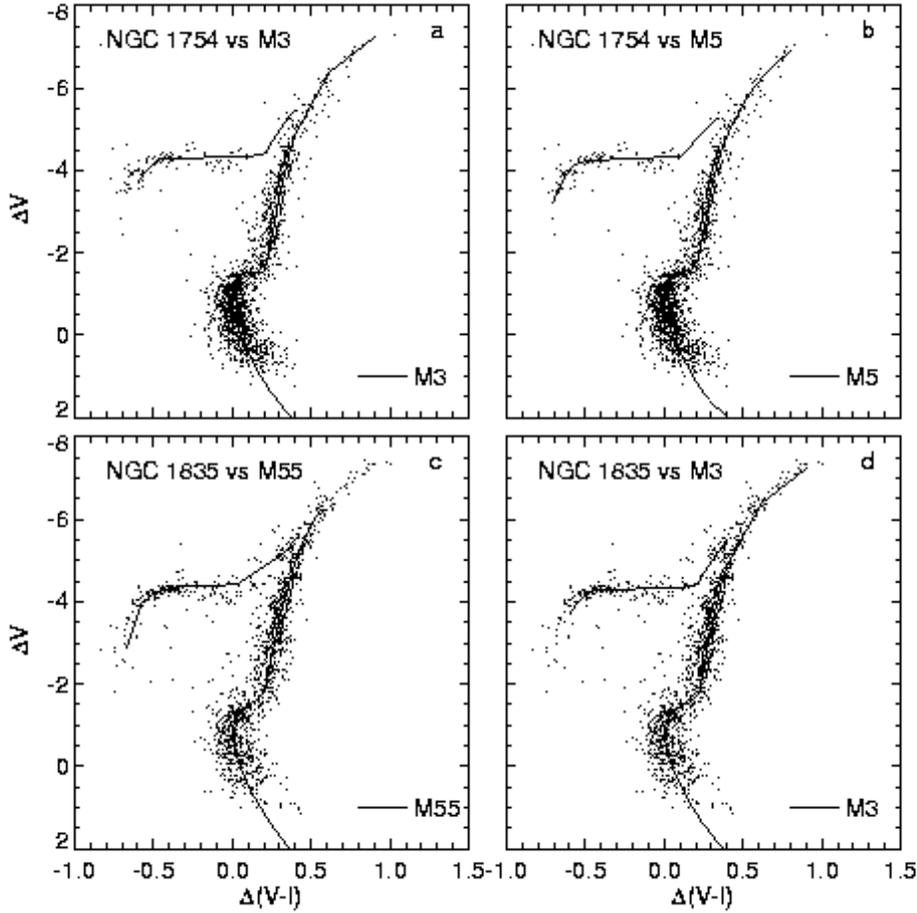}
\vspace{0.5cm}
\caption{The colour-magnitude diagrams of the LMC clusters shifted to match Milky Way globular cluster fiducial sequences in turnoff colour and HB magnitude.  The dashed line to the blue of the fiducial RGB shows the location it would have if the fiducial were 2 Gyr older, while the line to the red shows the RGB of a cluster 2 Gyr younger than the fiducial.  The locations of these lines were calculated from the models of VandenBerg (1997).}
\end{figure*}

\addtocounter{figure}{-1}
\begin{figure*}
\plotone{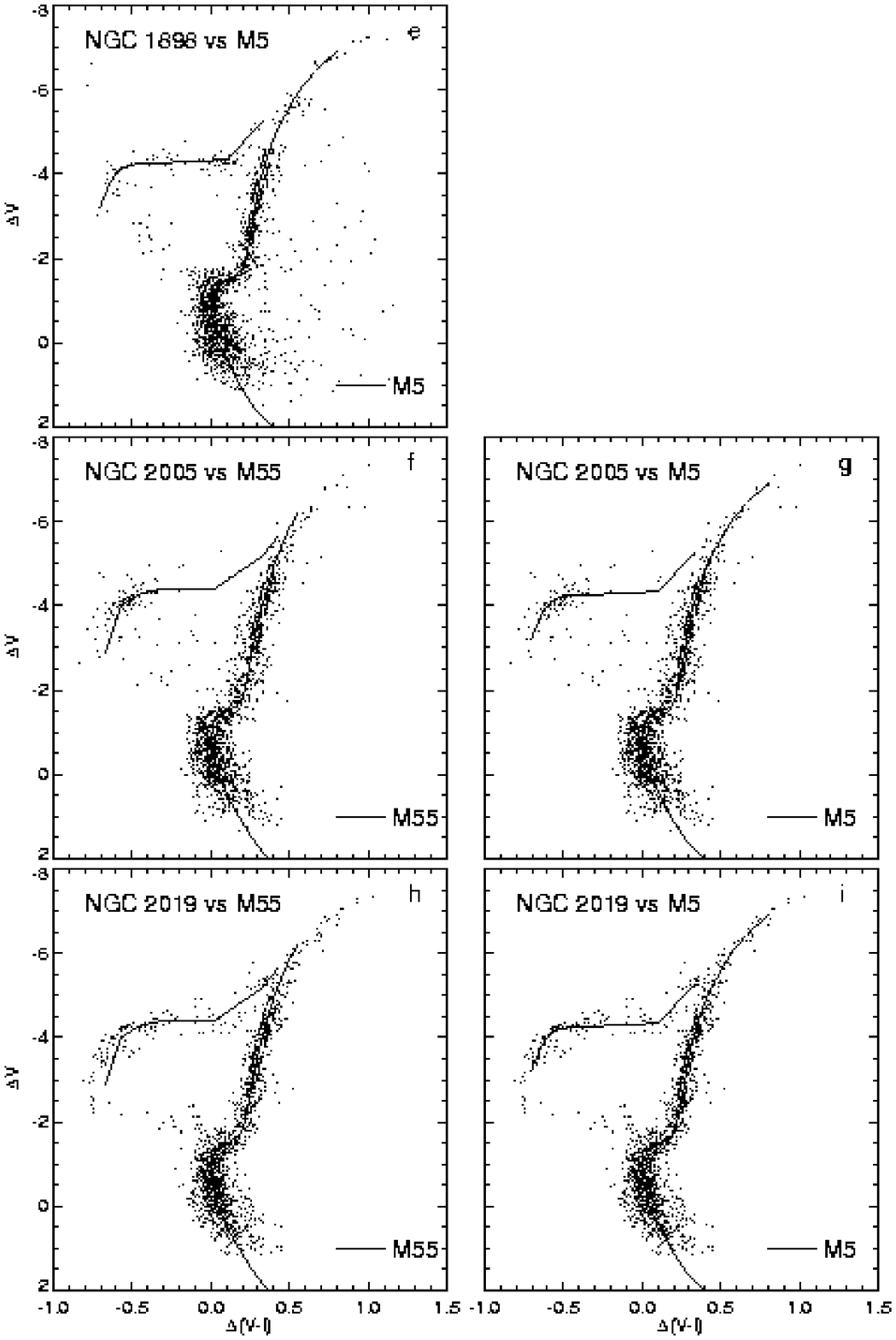}
\vspace{0.5cm}
\caption{cont.}
\end{figure*}

\subsection{Comparison of LMC and Milky Way globular cluster systems}
We used our knowledge of the abundances of the LMC clusters in conjunction with available methods to measure the ages of the clusters.  Due to the strong dependence on the uncertain photometric zero points, we avoided measuring the ages from a direct comparison of the CMDs to model isochrones.  Instead, we relied on differential techniques which are free of zero point dependence.  We used the approach of VBS to compare our CMDs to fiducial sequences of Milky Way globular clusters of similar abundance, measuring relative ages with an accuracy of $< \pm$1 Gyr from the difference in position of the RGBs.  We also measured $\Delta V{\rm ^{TO}_{HB}}$, which is a calibrated function of age.  While less accurate than the ``horizontal" VBS method, $\Delta V{\rm ^{TO}_{HB}}$ has the advantage of being fundamentally better understood.  Finally, as illustrated by LDZ, comparison of cluster HB morphologies with HB evolutionary tracks is a potentially very powerful way to establish the chronology of globular cluster formation.  While it is controversial, we used this technique to attempt to build a consistent picture of the LMC cluster ages.  Where inconsistencies arise, we can then interpret them to be due either to errors in the observations or a failure in some portion of the theoretical framework.

Each of the above age-dating techniques depends on the assumed metallicity, the VBS technique through the need to compare clusters of similar metallicity, $\Delta V{\rm ^{TO}_{HB}}$ through the metallicity dependence of both the magnitude of the HB and the magnitude of the MSTO, and the LDZ technique through metallicity being the ``first parameter" affecting HB morphology.  While we have available directly measured spectroscopic abundances of the clusters (O91), these abundances are based on difficult observations of only 1-2 giant stars per cluster.  We therefore measured the abundance of each cluster from the CMD using the formalism of Sarajedini (1994, hereafter S94), described in detail below.  Because these CMD-based abundances are rooted in well-understood photometric data of a large number of stars, we regard them as more reliable than the currently available spectroscopic abundances.

\subsubsection{Abundances and reddenings from the CMDs}
We followed S94 to derive abundances from the height of the RGB above the level of the HB; this method also yields the reddening from the colour of the RGB at the level of the HB.  As suggested in S94, we fit the portion of the RGB brighter than the HB with a 2nd order polynomial, which let us solve for $E(B-V)$ and [Fe/H] analytically.  We determined the level of the HB by taking the median of the points on the flat portion of the HB.  For NGC 2005, which has mostly blue HB stars, this involves some guesswork, so its parameters are likely more uncertain than those of the other clusters.

To calculate the uncertainties, we performed the polynomial fits to the RGB on a set of 100 Monte Carlo realizations of each CMD.  These simulated CMDs were constructed by choosing the appropriate number of stars from the luminosity functions of Vandenberg (1997) isochrones with abundances and reddenings approximately matching the observed CMDs.  We simulated the observational errors by applying the $V-I$ and $V$ shifts calculated from our artificial star database.  
After introducing a 0.1 magnitude Gaussian-distributed error in the $V$ magnitude of the HB, which was not simulated, we then calculated $E(B-V)$ and [Fe/H] for the simulated CMDs and compared the values to the input values.  We found typical errors of 0.15 dex in [Fe/H] and 0.01 magnitudes in $E(B-V)$ with systematic biases of up to 0.1 dex in [Fe/H] and 0.025 magnitudes in $E(B-V)$.  These biases are probably due to small scale changes in the photometry introduced by crowding, as illustrated in the example of Fig. 10.  Table 6 contains the calculated $E(B-V)$ and [Fe/H] values for the clusters, after correction for the biases found in the Monte Carlo experiments.

\subsubsection{Relative ages through comparison with Milky Way clusters}
We compared our LMC cluster CMDs to fiducial sequences of M3, M5, M13, M92 (Johnson \& Bolte 1997), and M55 (Mandushev et al. 1996 and personal communication), which span the range of metallicity $-1.4 \la$ [Fe/H] $\la -2.2$.  For consistency, we adopt the Zinn \& West (1984) abundances for all of the Milky Way clusters.  We used the technique described by VBS to establish the comparison between the LMC clusters and the Milky Way fiducials.  With this technique, the cluster CMD and a fiducial of similar metallicity are registered in the C-M plane through the colour of the main sequence turnoff (MSTO) and the magnitude of the point 0.05 magnitudes redder than the MSTO.  The difference in the lengths of the subgiant branches can then be easily measured and converted to a relative age through a calibration from model isochrones.

As an alternative to the process described above, which is fairly automatic
and unbiased, we also attempted to derive comparisons by eye, involving
both the disadvantage of possible subconscious bias and the advantage of
judgement based on knowledge. Each of the six LMC cluster CMDs was compared
to the fiducials of M3 and M13 and displacements in $V-I$ and in $V$ were
recorded for the best eye fits. (This was done by an author who had not, up
to that time, been involved with the initial fitting process).
Interestingly, the results were quite similar to those from the automatic
comparison. The average difference in the derived 
$V-I$ was 0.01 magnitudes, while that in $V$ was 0.11 mag. These by-eye
measurements confirm that the automatic
method had not introduced any unrealistic data into the problem.

\begin{figure}
\epsscale{1.0}
\plotone{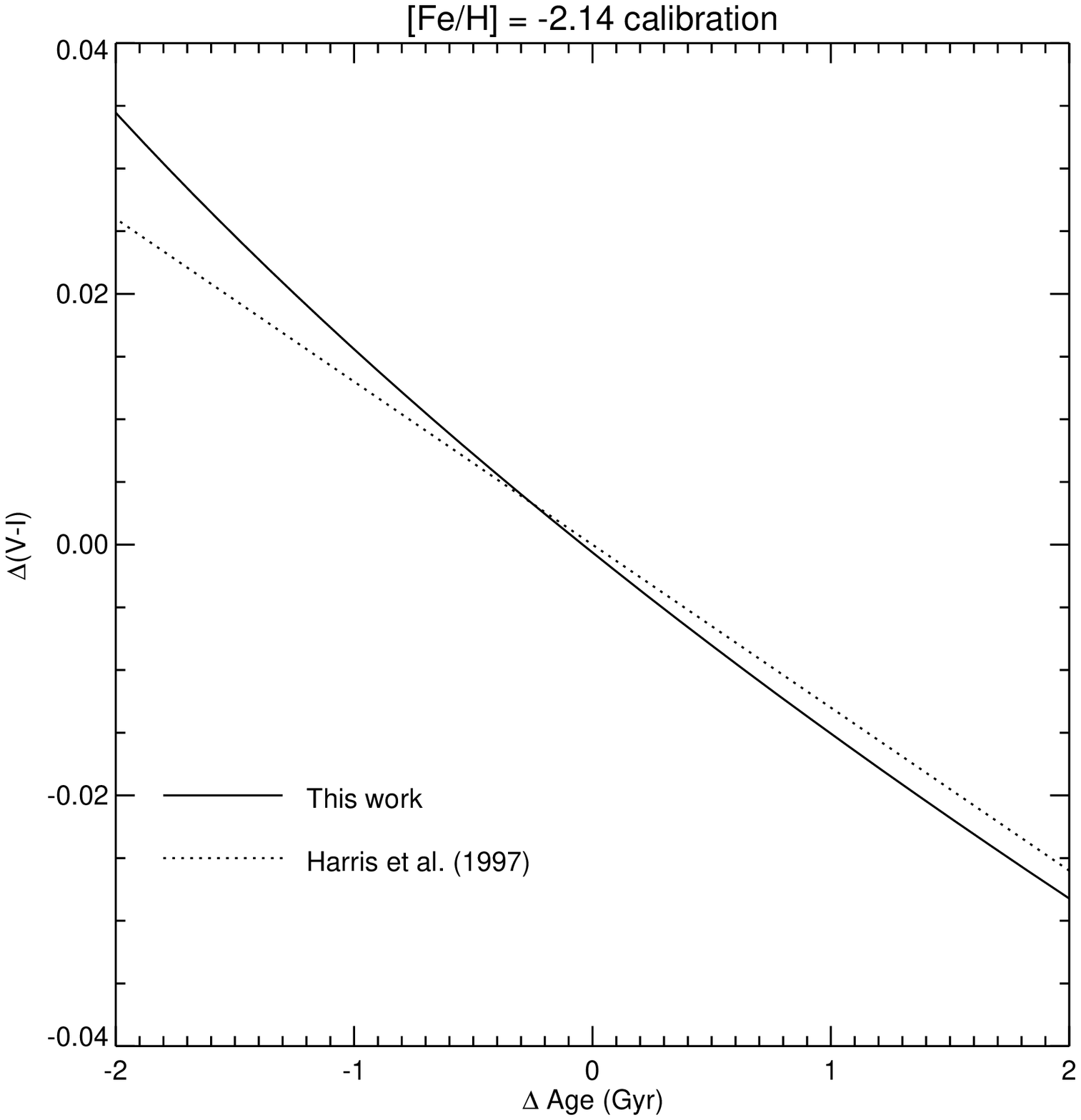}
\caption{Calibration of age difference as a function of colour offset of the RGB compared to the calibration of Harris et al. (1997) for the example of [Fe/H]$=-2.14$.  The calibration differences are small compared to the precision of the age measurements.}
\end{figure}

\subsubsection{Registration of the CMDs}
We measured the colour of the MSTO, $(V-I)_{\rm MSTO}$, of our clusters and the fiducials by fitting a second order
polynomial to a region around the turnoff and calculating the bluest point of the fit.  We iterated the fit until the $V-I$ colour of the bluest point of the polynomial fit changed by less than 5$\times 10^{-4}$ magnitudes, each time rejecting outlying points.  To measure the $V$ magnitude of the point 0.05 magnitudes redder than the MSTO ($V_{+0.05}$) we used a similar fitting procedure.  We fit a straight line to points near $V_{+0.05}$, rejecting outliers and iterating the fit until $V_{+0.05}$ changes by less than 0.01 magnitudes.  However, we found that after using $V_{+0.05}$ to register the LMC cluster CMDs with the Milky Way fiducials, the magnitudes of the HBs occasionally disagreed by $\ga$ 0.1 magnitudes, even for clusters of the same metallicity.  In these cases, we consider our eye estimates of the level of the HBs more reliable than the measured $V_{+0.05}$ points, as these $V_{+0.05}$ measurements are likely affected by incompleteness in the photometry.

Figs. 14a-i show the comparison of our LMC clusters with Milky Way fiducials after registration.  The dashed lines indicate the differences in position of the RGB that would be expected for a 2 Gyr age difference.  We only show those comparisons for which the abundances of the fiducials most closely match either the spectroscopic or CMD-based abundances.  For NGC 1754, NGC 1835, and NGC 1898, the difference between the comparisons based on the spectroscopic abundances and those based on the CMD abundances are small.  For NGC 2005 and NGC 2019, the CMD-based abundances are $\sim$0.6 dex higher than the spectroscopic ones.  The high CMD-based abundance of NGC 2005 especially comes as a surprise because the cluster has a very blue HB morphology.  While the discrepancy with the spectroscopic abundances is a cause for some concern, we note that the spectroscopic abundances of NGC 2005 and NGC 2019 are based solely on measurements of a single star in each cluster, which could be in error.  In addition, the CMD-based abundances produce better matches of the Galactic fiducials to the overall CMDs, reinforcing our greater confidence in these abundances.

\begin{table}
\caption{Calibration of Relative Ages}
\begin{tabular}{lccc}
[Fe/H] & 
$a$ &
$b$ &
$c$ \\
\hline
-1.14 & $8.51\times 10^{-4}$ & $9.70\times 10^{-3}$ & $-5.25\times 10^{-4}$ \\
-1.31 & $7.80\times 10^{-4}$ & $1.02\times 10^{-2}$ & $-6.11\times 10^{-4}$ \\
-1.41 & $7.91\times 10^{-4}$ & $1.05\times 10^{-2}$ & $-6.59\times 10^{-4}$ \\
-1.54 & $9.71\times 10^{-4}$ & $1.11\times 10^{-2}$ & $-7.49\times 10^{-4}$ \\
-1.61 & $1.12\times 10^{-3}$ & $1.14\times 10^{-2}$ & $-7.95\times 10^{-4}$ \\
-1.71 & $1.44\times 10^{-3}$ & $1.19\times 10^{-2}$ & $-8.87\times 10^{-4}$ \\
-1.84 & $2.07\times 10^{-3}$ & $1.26\times 10^{-2}$ & $-9.68\times 10^{-4}$ \\
-2.01 & $2.51\times 10^{-3}$ & $1.36\times 10^{-2}$ & $-1.16\times 10^{-3}$ \\
-2.14 & $2.81\times 10^{-3}$ & $1.45\times 10^{-2}$ & $-1.28\times 10^{-3}$ \\
\hline
\end{tabular}
\end{table}

\subsubsection{Relative ages and age errors}
To measure the difference in position of the RGBs in Figs. 14a-i, 
we fit the piece of the fiducial extending from $-5 \le \Delta V \le -2$ to the CMD points in the same $\Delta V$ range.  To map the differences in RGB position to age differences between the Galactic and LMC clusters, we used a calibration based on the new isochrones of VandenBerg (1997).  These isochrones are calculated for more than a dozen metallicities in the range $-2.3 \le$ [Fe/H] $\le -0.3$ for ages of 8-18 Gyr in 2-Gyr steps and fit observations of M92 extremely well.  After registering the six isochrones of a given metallicity in the $\Delta (V-I),\Delta V$ plane, we fit a piece of the 14 Gyr isochrone spanning the range $-5 \le \Delta V \le -2$ to similar pieces of the other five isochrones.  To establish the calibration over a continuum of age differences, we fit a 2nd order polynomial to 
the offsets needed to shift this portion of the RGB of the 14 Gyr isochrone to the RGBs of the other isochrones.  The choice of the 14 Gyr isochrone as the fiducial for this calibration is inconsequential, as the change of $V-I$ colour of the RGB is nearly linear with age for ages between 12 and 18 Gyr, the likely age range of old globular clusters.  Table 5 lists the polynomial coefficients of the fits to the equation $\Delta {\rm Age}=a+b\Delta (V-I)+c\Delta (V-I)^2$ for the range of metallicities considered in this paper.  Fig. 15 shows our calibration for [Fe/H]=$-2.14$ compared to that of Harris et al. (1997).  Over a range of age differences of $\pm$1 Gyr, our calibration differs by $\sim$15 per cent from that of Harris et al. (1997).  This difference may be attributed to the fact that we used a large portion of the RGB to measure the age differences while Harris et al. (1997) apparently used only the difference in RGB color at $\Delta V=2.5$.  Because we can more accurately measure the differences in RGB colors if we use many points on the RGB, we chose to use our calibration over that of Harris et al. (1997).  However, as the difference in the calibration will only produce differences in the relative age measurements of $\la$0.2 Gyr, the choice has little consequence for our results.

\begin{figure*}
\epsscale{0.95}
\plotone{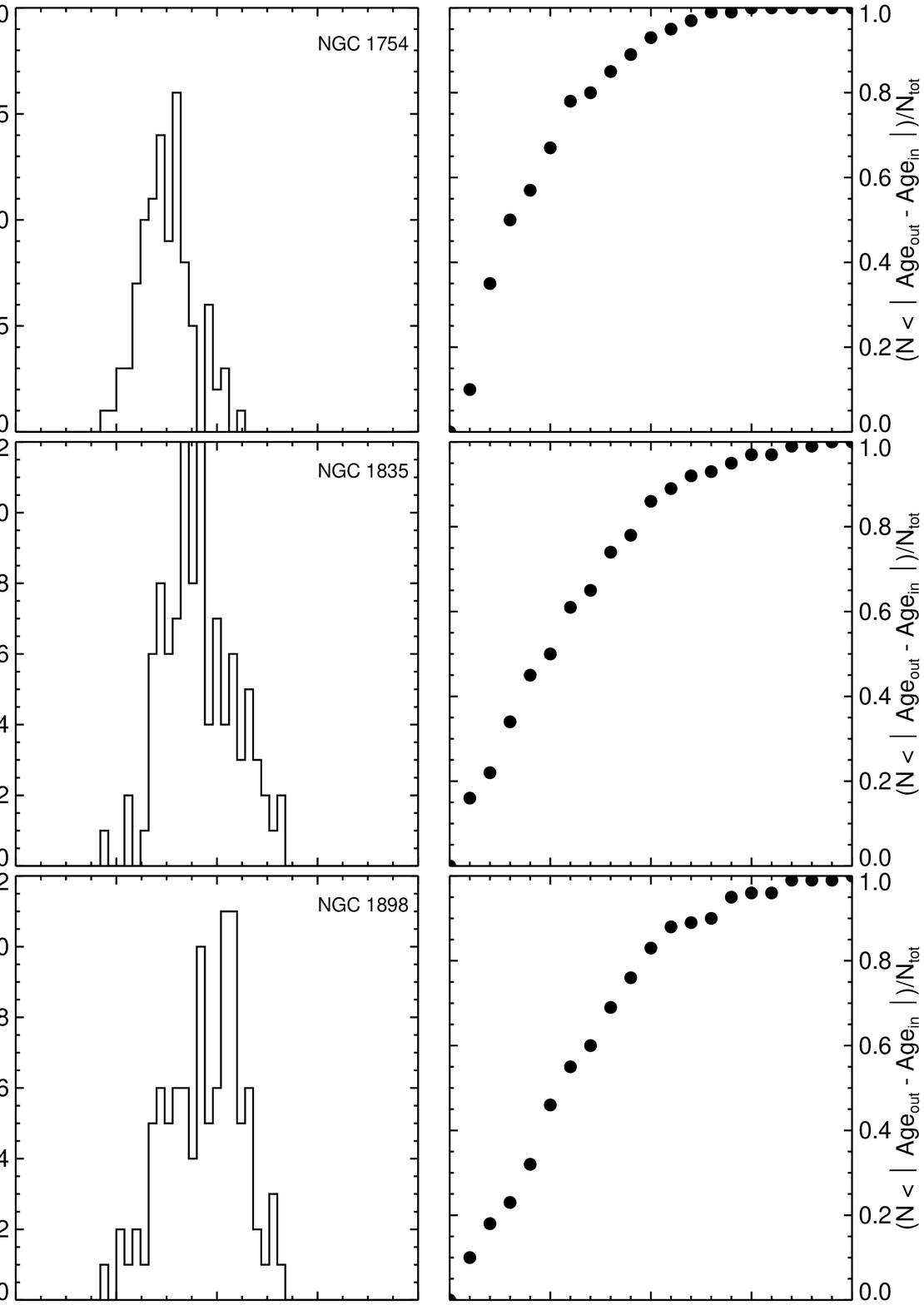}
\hspace{1.cm}
\plotone{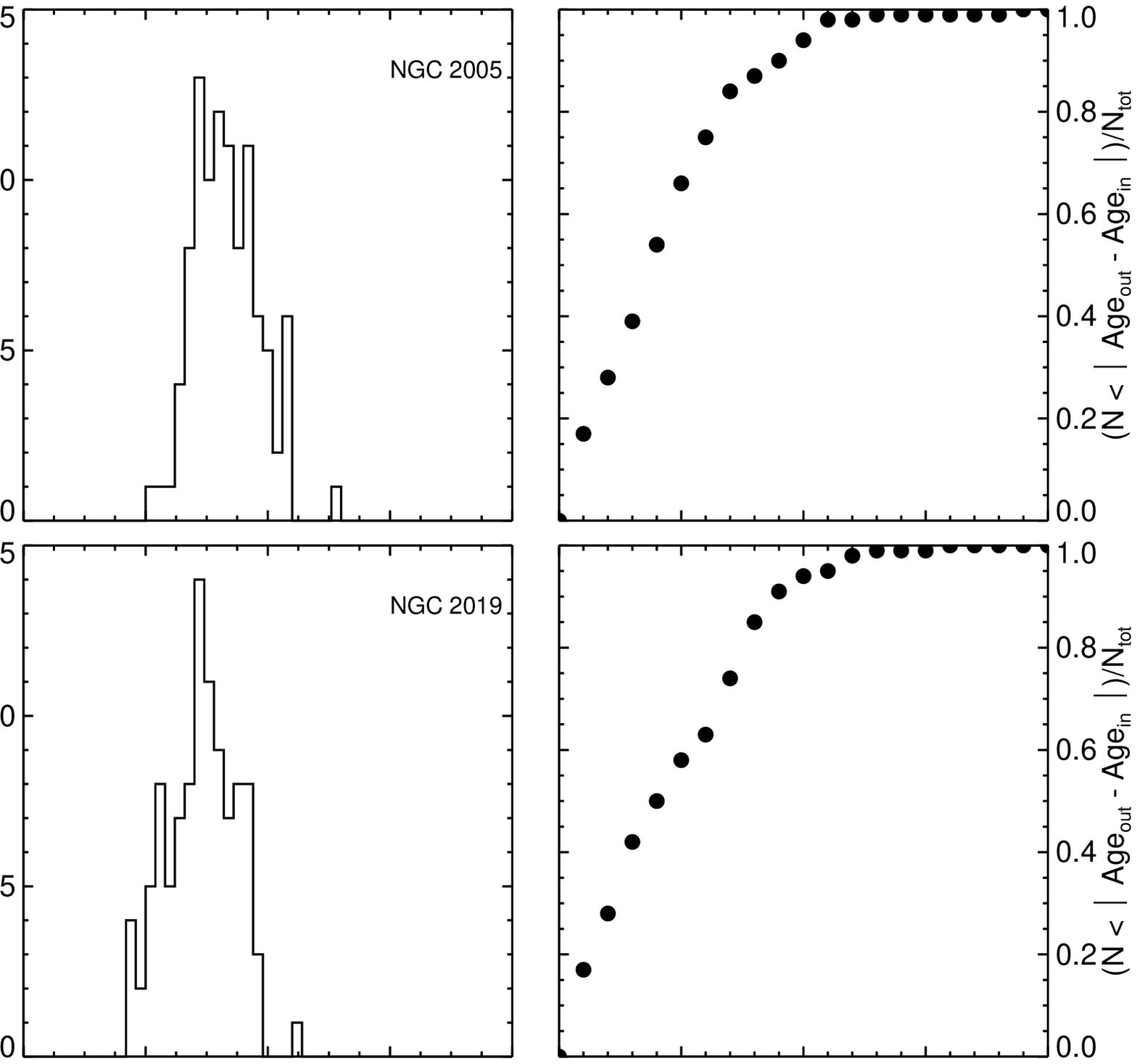}
\vspace{0.2cm}
\caption{Plots of the distribution of errors in the relative age measurement and cumulative distribution of errors around the mean calculated from the Monte Carlo experiments described in the text.  70 per cent of the cumulative distributions 
are enclosed within $\la \pm 1$ Gyr.  However, each cluster shows a bias of 0-1 
Gyr towards measuring younger relative ages.}
\end{figure*}

\begin{table*}
\caption{Derived Parameters of LMC Globular Clusters}
\begin{tabular}{llcccccc}
& & NGC 1754 & NGC 1835
& NGC 1898 & NGC 2005 & NGC 2019 \\
&  \\
\hline
Cluster center\dotfill & RA (2000.0) & 4$^{\rm h}$54$^{\rm m}$18\fs70 & 5$^{\rm h}$05$^{\rm m}$6\fs44 & 5$^{\rm h}$16$^{\rm m}$42\fs41 & 5$^{\rm h}$30$^{\rm m}$10\fs36 & 5$^{\rm h}$31$^{\rm m}$56\fs73 \\
& Dec & -70\fdg26\arcmin32\farcs1 & -69\fdg24\arcmin14\farcs5 & -69\fdg39\arcmin24\farcs6 & -69\fdg45\arcmin9\farcs0 & -70\fdg09\arcmin33\farcs3 \\
Structural & $r_c$ (\arcsec) & 3.6 & 2.7 & 5.2 & 1.3 & 0.9 \\
parameters\dotfill & $\log_{10}(r_t/r_c)$ & 1.5 & 1.5 & 1.9 & 1.9 & 2.2 \\
$[\rm{Fe/H}]$\dotfill & O91 & -1.54 $\pm$ 0.2 & -1.79 $\pm$ 0.2 & -1.37 $\pm$ 0.2 & -1.92 $\pm$ 0.2 & -1.81 $\pm$ 0.2 \\
& S94 method & -1.42 $\pm$ 0.15 & -1.62 $\pm$ 0.15 & -1.18 $\pm$ 0.16 & -1.35 $\pm$ 0.16 & -1.23 $\pm$ 0.15 \\
E(B-V)\dotfill & O91 & 0.08 $\pm$0.02 & 0.13 $\pm$ 0.02 & 0.06 $\pm$ 0.02 & 0.12 $\pm$ 0.02 & 0.12 $\pm$ 0.02 \\
& MW GC comparison  \\
& \hspace{0.1in}O91 abundance & 0.10 $\pm$0.02 & 0.07 $\pm$ 0.02 & 0.08 $\pm$ 0.02 & 0.12$\pm$ 0.02 & 0.07 $\pm$ 0.02  \\
& \hspace{0.1in}S94 abundance & 0.09 $\pm$0.02 & 0.12 $\pm$0.02 & 0.08 $\pm$0.02 & 0.09 $\pm$0.02 & 0.05 $\pm$0.02 \\
& S94 method & 0.082 $\pm$0.01 & 0.036 $\pm$ 0.01 & 0.046 $\pm$ 0.01 & 0.068 $\pm$ 0.01 & 0.034 $\pm$ 0.01 \\
& adopted & 0.09 $\pm$0.02 & 0.08 $\pm$0.02 & 0.07 $\pm$0.02 & 0.10 $\pm$0.02 & 0.06 $\pm$0.02 \\
$V_{\rm HB}$\dotfill & & 19.57 $\pm$ 0.10 & 19.30 $\pm$ 0.10 & 19.41 $\pm$ 0.10 & 19.39 $\pm$ 0.10 & 19.24 $\pm$ 0.10 \\
$(V-I)_{\rm MSTO}$\dotfill & & 0.72 $\pm$ 0.01 & 0.66 $\pm$ 0.02 & 0.69 $\pm$ 0.01 & 0.71 $\pm$ 0.01 & 0.68 $\pm$ 0.02 \\
(B-R)/(B+V+R)\dotfill & & 0.47 $\pm$ 0.07 & 0.48 $\pm$ 0.05 & -0.08 $\pm$ 0.10 & 0.87 $\pm$ 0.04 & 0.56 $\pm$ 0.07 \\
\underline{Ages} \\
Age (LMC -- MW cluster)\dotfill & O91 abundance  \\
& ~~MW cluster: M3 & 2.56 $\pm$ 0.53 & $-$ & $-$ & $-$ & $-$  \\
& ~~MW cluster: M5 & $-$ & $-$ & -0.45 $\pm$ 0.81 & $-$ & $-$  \\
& ~~MW cluster: M55 & $-$ & 1.03 $\pm$ 0.76 & $-$ & 0.35 $\pm$ 0.54 & 0.35 $\pm$ 0.66 \\
& S94 abundance  \\
& ~~MW cluster: M3 & $-$ & 2.23 $\pm$ 0.76 & $-$ & $-$ & $-$  \\
& ~~MW cluster: M5 & 2.42 $\pm$ 0.53 & $-$ & -0.45 $\pm$ 0.81 & 0.42 $\pm$ 0.54 & 0.53 $\pm$ 0.66 \\
$\Delta V{\rm ^{TO}_{HB}}$\dotfill & & 3.51 $\pm$ 0.13 & 3.54 $\pm$ 0.16 & 3.44 $\pm$ 0.15 & 3.51 $\pm$ 0.29 & 3.59 $\pm$ 0.17 \\
Age($\Delta V{\rm ^{TO}_{HB}}$)\dotfill & O91 abundance & 15.6 $\pm$ 2.3 & 16.6 $\pm$ 2.9 & 14.0 $\pm$ 2.3 & 16.6 $\pm$ 5.1 & 17.8 $\pm$ 3.2  \\
& S94 abundance & 15.6 $\pm$ 2.2 & 16.2 $\pm$ 2.8 & 13.5 $\pm$ 2.2 & 15.5 $\pm$ 4.9 & 16.3 $\pm$ 3.1 \\
\underline{Distances} \\
$(m-M)_V$\dotfill & MW GC comparison  \\
& ~~O91 abundance & 19.00 $\pm$ 0.15 & 18.67 $\pm $ 0.15 & 18.69 $\pm $ 0.15 & 18.71 $\pm$ 0.15 & 18.64 $\pm$ 0.15  \\
& ~~S94 abundance & 18.87 $\pm$ 0.15 & 18.77 $\pm $ 0.15 & 18.69 $\pm $ 0.15 & 18.69 $\pm$ 0.15 & 18.62 $\pm$ 0.15 \\
$(m-M)_\circ$\dotfill & MW GC comparison  \\
& ~~O91 abundance & 18.70 $\pm$ 0.16 & 18.46 $\pm$ 0.16 & 18.45 $\pm$ 0.16 & 18.31 $\pm$ 0.16 & 18.38 $\pm$ 0.16  \\
& ~~S94 abundance & 18.60 $\pm$ 0.16 & 18.65 $\pm$ 0.16 & 18.45 $\pm$ 0.16 & 18.40 $\pm$ 0.16 & 18.46 $\pm$ 0.16 \\
$(m-M)_V$\dotfill & $V_{\rm HB}$  \\
& ~~O91 abundance & 18.90 $\pm$ 0.11 & 18.68 $\pm $ 0.11 & 18.70 $\pm $ 0.11 & 18.79 $\pm$ 0.11 & 18.62 $\pm$ 0.11  \\
& ~~S94 abundance & 18.87 $\pm$ 0.10 & 18.64 $\pm $ 0.10 & 18.67 $\pm $ 0.10 & 18.68 $\pm$ 0.10 & 18.51 $\pm$ 0.10 \\
$(m-M)_\circ$\dotfill & $V_{\rm HB}$  \\
& ~~O91 abundance & 18.62 $\pm$ 0.12 & 18.43 $\pm $ 0.12 & 18.49 $\pm $ 0.12 & 18.48 $\pm$ 0.12 & 18.44 $\pm$ 0.12  \\
& ~~S94 abundance & 18.60 $\pm$ 0.12 & 18.40 $\pm $ 0.12 & 18.45 $\pm $ 0.12 & 18.37 $\pm$ 0.12 & 18.32 $\pm$ 0.12 \\
\hline
\end{tabular}
\end{table*}

Table 6 contains the ages we derive, in Gyr, for the LMC clusters relative to the appropriate comparison clusters.  On average, the LMC clusters have the same ages as the Milky Way clusters to within 1.0 $\pm$ 1.3 Gyr, reinforcing our earlier statement that the LMC old globular clusters are very similar to those in the Milky Way.  In order to examine this global statement on a cluster-to-cluster basis, however, we need to establish the accuracy of the individual relative age measurements.  We did this by performing the relative age measurement on the sets of 100 Monte Carlo realizations described previously.  For each simulated CMD, we measured $(V-I)_{\rm MSTO}$ and $V_{+0.05}$ as for the observed CMD and registered the simulated CMD and the error-free input isochrone in the $\Delta (V-I),\Delta V$ plane.  The simulated measurements of $V_{+0.05}$ confirm that low completeness affects this measurement, justifying our use of the HB $V$ magnitude to perform the vertical registration.  Since we do not simulate the HB stars, we assumed that we can correctly perform the vertical registration to within 0.1 $V$ magnitudes.  Thus, we vertically registered the simulated CMD according to the exact $V_{+0.05}$ value calculated from the isochrone, but applied a random offset selected from a Gaussian with a 1-$\sigma$ dispersion of 0.1 magnitudes.  We then measured the age difference between the simulated CMD and the input isochrone as we did for the real clusters.  Figs. 16a-e show the resulting distributions of age differences and the cumulative distributions of absolute age differences around the mean.  The distributions of age differences are narrow, with 70 per cent of the distributions enclosed within 0.7-0.8 Gyr of the mean.  However, three of the clusters, NGC 1754, NGC 2005, and NGC 2019, show systematic biases towards recovering ages $\sim$1 Gyr younger than the age of the input isochrone, while NGC 1835 has a bias towards recovering ages $\sim$0.5 Gyr younger than the input.
These biases are likely due to the crowding errors in the photometry illustrated by the example of Fig. 10.  Although ignoring the biases would not significantly affect our results, we have corrected for them in the relative age measurements reported in Table 6.

\subsubsection{Ages from $\Delta V{\rm ^{TO}_{HB}}$}
To measure $\Delta V{\rm ^{TO}_{HB}}$ we used the values of $V_{\rm HB}$ measured in section 6.2.1 and the magnitude of the MSTO ($V_{\rm MSTO}$) implied by the polynomial fit to the points near the turnoff.  Because the MSTO region is nearly vertical in the CMD and the measurement of $V_{\rm HB}$ is subject to some interpretation, the error in  $\Delta V{\rm ^{TO}_{HB}}$ may be larger than reported here.  We estimate the error in $V_{\rm HB}$ to be $\pm$0.1 magnitudes, while for the error in $V_{\rm MSTO}$ we adopt the error calculated from the covariance matrix of the polynomial fit.

To convert $\Delta V{\rm ^{TO}_{HB}}$ to age, we adopted the calibration of Chaboyer, Demarque, \& Sarajedini (1996), which is based on a recent set of Yale isochrones (Chaboyer \& Kim 1995).  The calibration depends on an assumed $M_V({\rm RR})-$[Fe/H] relation; their preferred relation is $M_V({\rm RR}) = 0.20{\rm [Fe/H]} + 0.98$.  Recent work based on {\it HIPPARCOS} parallaxes and proper motions of field RR Lyrae (Tsujimoto, Miyamoto, \& Yoshii 1998) finds a similar slope and zero point, $M_V({\rm RR}) = 0.20{\rm [Fe/H]} + 0.91$, albeit with large error.  The independent analysis of Fernley et al. (1998) gives $M_V({\rm RR}) = 0.18{\rm [Fe/H]} + 1.05$, which agrees, within the error, with the Tsujimoto et al. (1998) result.  As the Chaboyer et al. (1996) preferred zero point of the $M_V({\rm RR})-$[Fe/H] relation is the exact average of the two available {\it HIPPARCOS}-based zero points, we adopt their preferred relation.  An advantage of using this relation is that we can easily compare the ages we derive from $\Delta V{\rm ^{TO}_{HB}}$ with those of the Galactic globular clusters.  As has been noted, however, this relation disagrees strongly with the distance to the LMC implied by the first {\it HIPPARCOS} Cepheid work (Feast \& Catchpole 1997), but not with the recent {\it HIPPARCOS} Cepheid analysis of Madore \& Freedman (1998).  The $\Delta V{\rm ^{TO}_{HB}}$ ages we measure are listed in Table 6, for both the O91 and CMD-based abundances.

\begin{figure*}
\epsscale{1.0}
\plotone{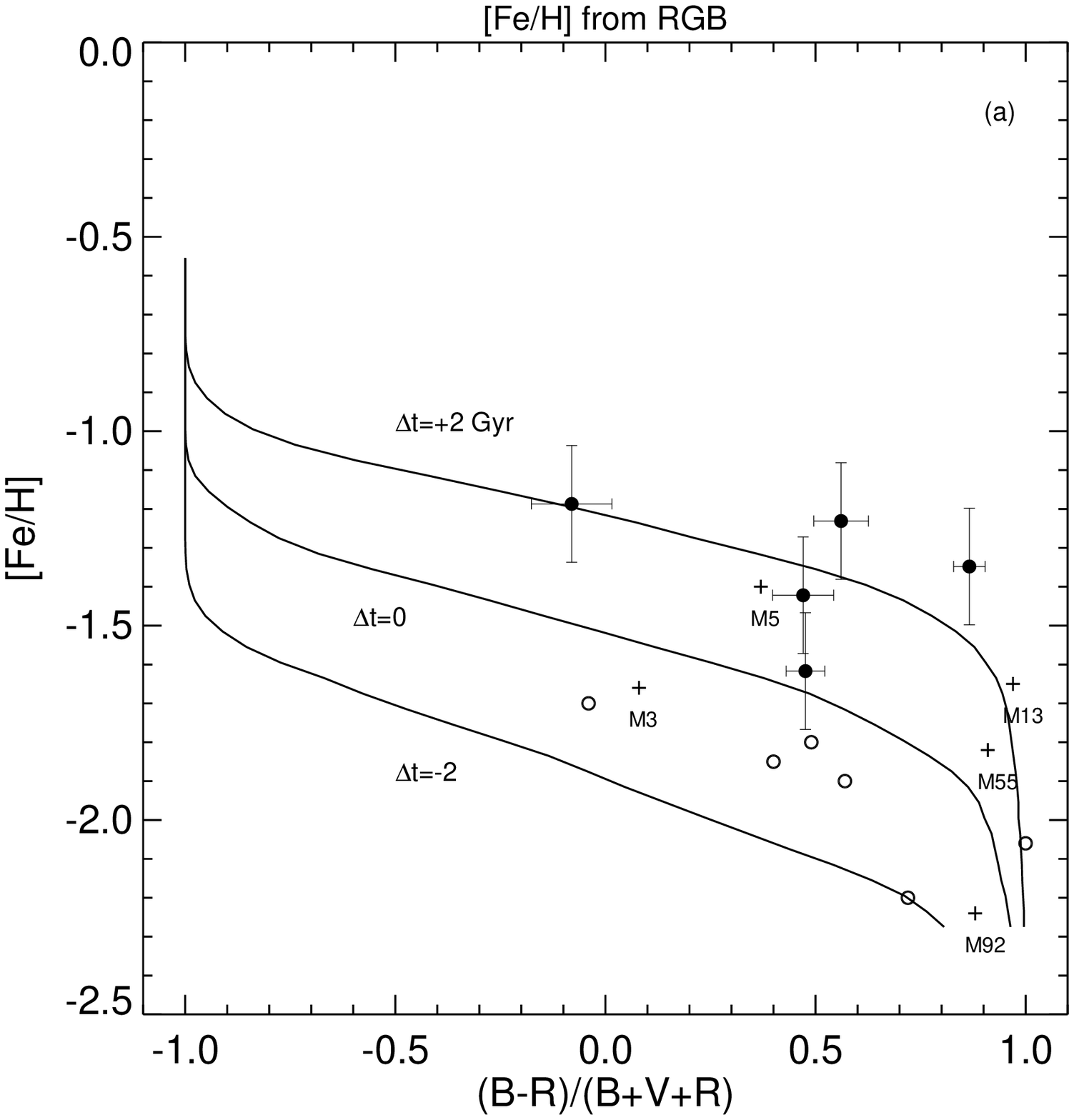}
\plotone{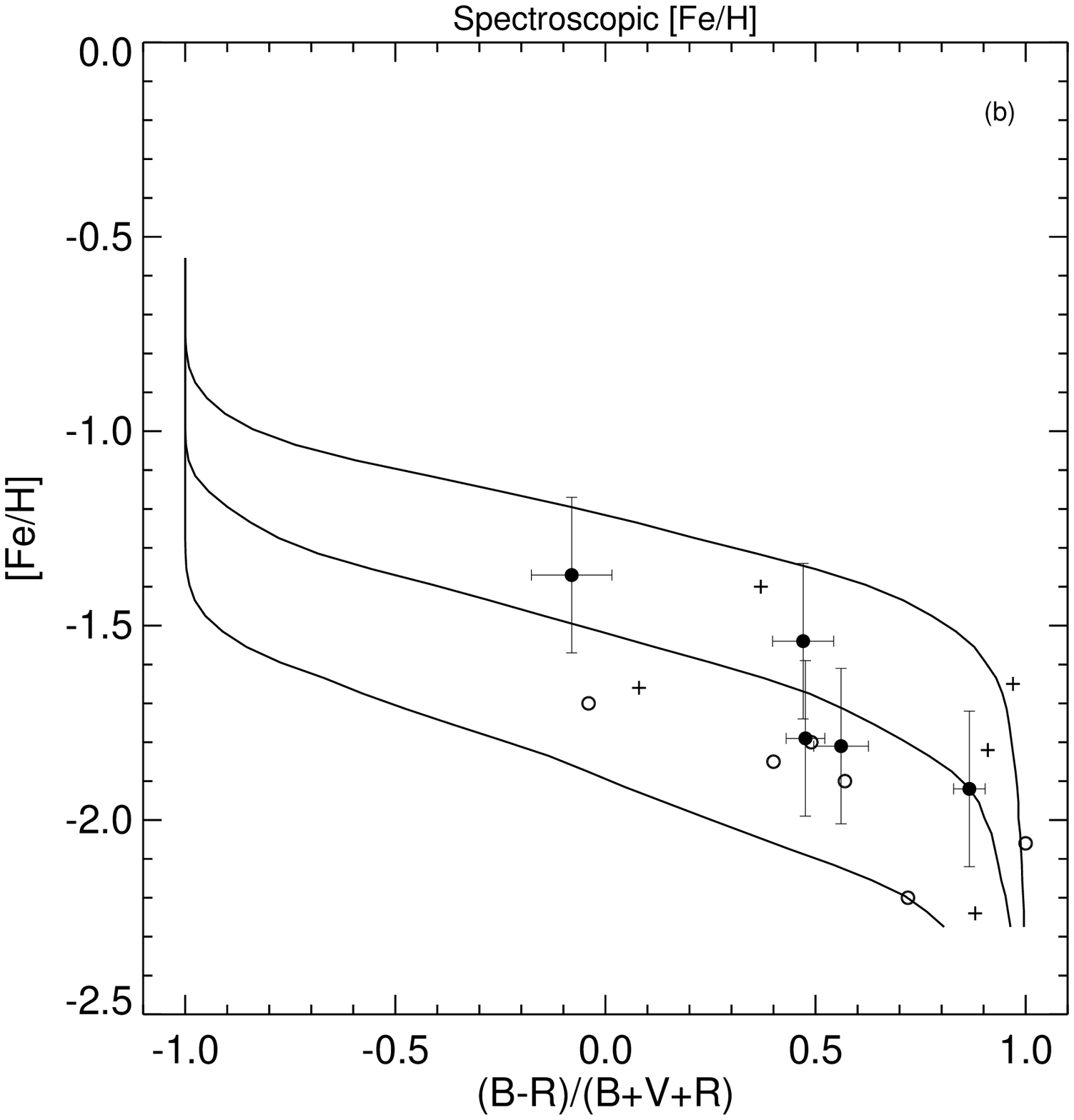}
\caption{Comparison of the horizontal branch morphologies of the LMC clusters of this study (filled circles), LMC clusters from Walker (1992b) (open circles), and Milky Way comparison clusters used in this study (crosses) using the the CMD-based abundances ($a$) and the spectroscopic abundances of Olszewski et al. (1991) ($b$).  The solid lines are HB evolutionary tracks from Lee et al. (1994) showing age differences of $\pm$2 Gyr.}
\end{figure*}

\subsubsection{Analysis of horizontal branch morphology}
An interesting question is whether the LMC clusters show a variation of HB morphology with metallicity similar to the Milky Way clusters.  We measured the HB morphology of our clusters through the commonly used (B-R)/(B+V+R) index, where B is the number of stars to the blue of the instability strip, R the number of stars to the red of the instability strip, and V the number RR Lyrae variables.  Because we do not have sufficient time resolution in our images to identify RR Lyrae variables from their light curves, we rely on the approximate location of the instability strip in the CMD plane to determine the boundaries of the B, V, and R zones.  Our chosen boundaries are $V-I$=0.23 for the blue edge of the instability strip and $V-I$=0.57 for the red edge.  Table 6 contains our measured values of (B-R)/(B+V+R) for each cluster, corrected for incompleteness, along with uncertainties calculated from Poisson counting statistics and the uncertainties in the completeness corrections.

In Figs. 17a\&b we plot [Fe/H] vs. (B-R)/(B+V+R) for the LMC clusters studied here, for the outer LMC clusters using data from Walker (1992b), and for selected Milky Way clusters from Table 1 of LDZ.  We show separate plots for the O91 spectroscopic abundances and the CMD-based abundances.  Depending on which set of abundance measurements is adopted, we arrive at slightly different conclusions.  If we choose to adopt the spectroscopic abundances, we conclude that, based on the HB morphologies, our LMC clusters are similar to the ``younger halo" of the Milky Way.  On the other hand, if we adopt the CMD-based abundances, we find that the LMC clusters are similar to the oldest Milky Way globulars, with NGC 1835 perhaps being $\sim$2 Gyr younger.  As we are more confident in the CMD-based abundances, the HB models suggest that the LMC is as old as the Milky Way.
Independent of the abundances used, all of the clusters, with the possible exception of NGC 1835, fall on the same HB evolutionary sequence.  The similarity in age of the clusters implied by the HB sequences agrees with the narrow age spread implied by both the comparison with Milky Way fiducials and the measurements of $\Delta V{\rm ^{TO}_{HB}}$.  It is also interesting to note that none of the LMC clusters studied here bears similarity to Ruprecht 106 or Pal 12, which have been suggested to be captures from the Magellanic Clouds (Lin \& Richer 1992; cf. Rup 106 CMD of Buonanno et al. 1993).

\subsubsection{Summary of reddenings and distances}
We estimated the reddenings of the clusters in two ways.  First, the horizontal shifts needed to register our CMDs with the the Milky Way fiducials yields the reddening difference between the LMC clusters and the Milky Way comparison clusters.  As the reddenings of the Milky Way clusters is well known, we can calculate the reddenings of the LMC clusters.  Second, as previously discussed, we use the S94 method to measure the reddening from $(V-I)_{\rm g}$, the colour of the giant branch at the level of the HB.  Table 6 contains our measurements of $E(B-V)$ as well as the estimates of Suntzeff et al. (1992).  For the Milky Way comparison method, we report separate values for the two possible abundance systems.  In all calculations, we have used $E(B-V)=E(V-I)/1.3$ (Dean, Warren, \& Cousins 1978).  As the dispersion in the measurements is small, we adopt for the $E(B-V)$ of each cluster the average of our measurements, also listed in Table 6.

Our CMDs are not sufficiently deep to accurately measure the distances to the clusters directly from a fit to the unevolved main sequence.  However, we can use the vertical shifts measured in Section 6.2.3 to measure the distances to the LMC clusters based on adopted distances to the Milky Way comparison clusters.  In addition, by adopting an $M_V({\rm RR)-[Fe/H]}$ relation, we can measure the distances to the LMC clusters from $V_{\rm HB}$.  By demanding that the distances measured with our data should be consistent with those implied by observations of LMC field RR Lyrae, we can use the distance measurements to check the internal consistency of our measurements of the reddenings and abundances, in addition to establishing the relative distances between the clusters.  For our calculations, we adopt distances to M3, M5, and M55 that are consistent with the LMC modulus of 18.5 used in Walker's (1992c) analysis of the zero point of the $M_V({\rm RR})-$[Fe/H] relation; we use the preferred $M_V({\rm RR)-[Fe/H]}$ relation of Chaboyer et al. (1996), $M_V({\rm RR)=0.20[Fe/H]+0.98}$, which is consistent with the SN1987A distance to the LMC of 18.37 (Gould 1995).  While other analyses of the SN1987A ring find larger LMC distances (Panagia et al. 1991, Sonneborn et al. 1997), the exact distance is irrelevant for the consideration of the {\it internal} consistency of our measurements.

Fig. 18 shows our distance modulus measurements for both the Milky Way comparison and $V_{\rm HB}$ methods and for both the O91 and CMD-based abundance sources.  We compare our measurements with the distance modulus gradient across the LMC of Caldwell \& Coulson (1986).  In the top panels, (a) and (b), we have used the O91 abundances while in the lower panels, (c) and (d), we have used the CMD-based abundances.  The panels on the left, (a) and (c), show the results derived from the comparison with Milky Way clusters while the panels on the right, (b) and (d), show the results of adopting $M_V({\rm RR)=0.20[Fe/H]+0.98}$ in conjunction with $V_{\rm HB}$.  We have subtracted an LMC distance modulus of 18.5 from the left panels and 18.37 from the right panels.  While the errors are large, the plots suggest that the CMD-based abundances produce a greater consistency between the cluster distances and the LMC modulus.  All of the distances, with the possible exception of NGC 1754, appear consistent with lying in the plane of the LMC.

\begin{figure}
\plotone{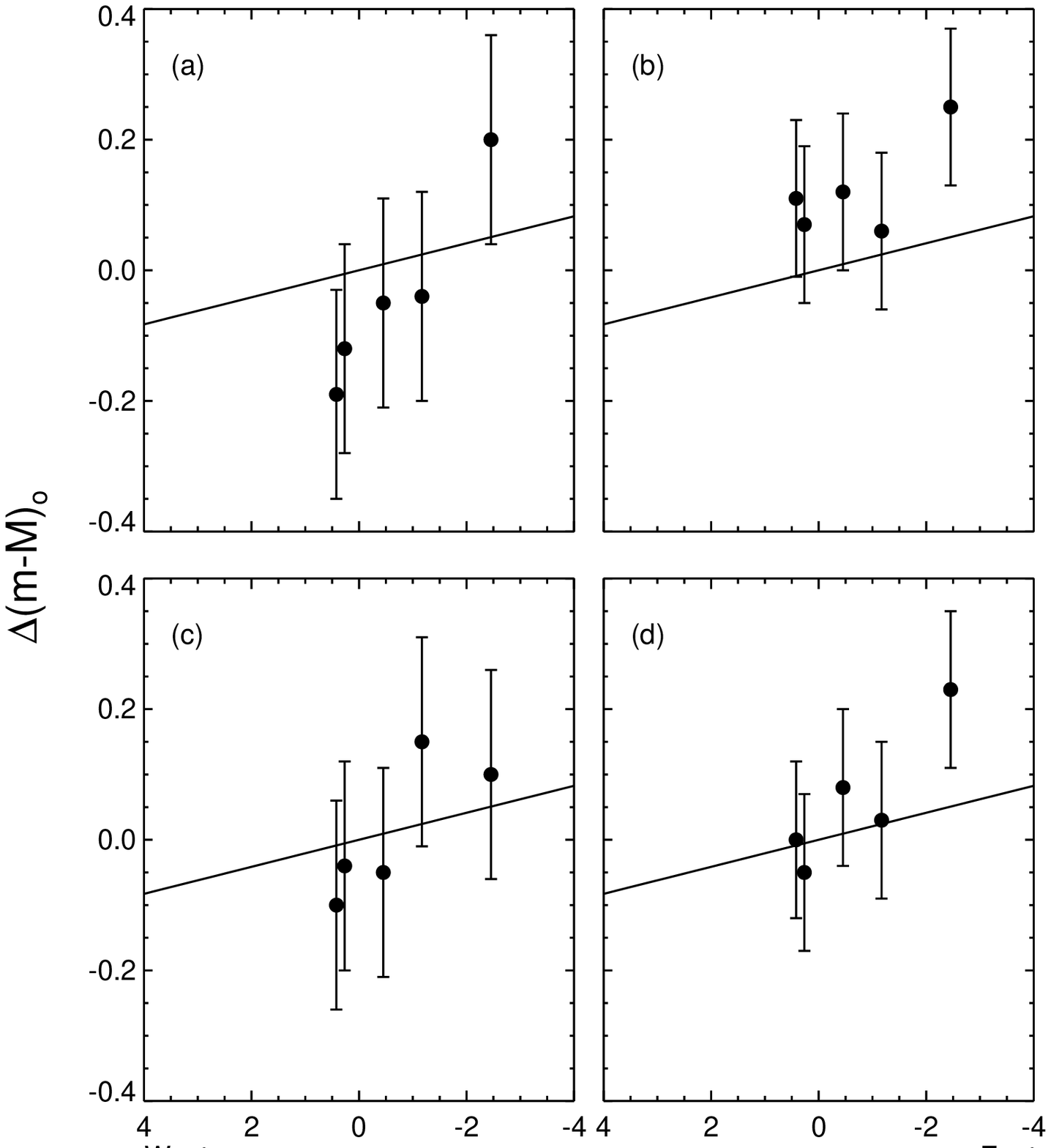}
\vspace{0.2CM}
\caption{Comparison of the distance moduli of the LMC clusters implied by two different methods using two sources of abundances with the model of the tilt of the plane of the LMC of Caldwell \& Coulson (1986).  ($a$) and ($b$) incorporate the spectroscopic abundances of Olszewski et al. (1991) while ($c$) and ($d$) use the CMD-based abundances.  ($a$) and ($c$) show the distances implied by the comparison to Milky Way clusters while ($b$) and ($d$) show the moduli implied by the horizontal branches, assuming $M_V({\rm RR})=0.20{\rm [Fe/H]}+0.98$ (Chaboyer et al. (1996).  A modulus of 18.5 has been subtracted from ($a$) and ($c$) and 18.37 from ($b$) and ($d$) to show the consistency with the distances implied by LMC RR Lyrae observations and the adopted zero points of the $M_V({\rm RR})-{\rm [Fe/H]}$ relation.}
\end{figure}

\section{Summary and Conclusions}
The main results of our analysis are that the six old LMC globular clusters NGC 1754, NGC 1835, NGC 1898, NGC 1916, NGC 2005, and NGC 2019 are clearly very similar in age, abundance, and HB morphology to the globular clusters of the Milky Way halo.  Excluding NGC 1916 from the bulk of the analysis because of the difficulty of treating its differential reddening, we have explored the similarity in age through the use of three age-dating techniques: comparison of our CMDs with Milky Way globular cluster fiducial sequences through the ``horizontal method" (VBS), measurement of $\Delta V{\rm ^{TO}_{HB}}$ (``vertical method"), and comparison of our HB data with HB evolutionary models from LDZ.  While showing the clear similarity in age with Milky Way globular clusters, each of these techniques additionally suggest that we detect no internal age spread in this set of LMC globular clusters.  We have measured the abundances of the clusters from the slope of the RGB (S94).  While the mean abundance that we derive is higher than that measured by O91, it is not grossly different.  We have argued, however, that the CMD-based abundances should be considered more reliable.

Throughout our analysis, we have tried to be as careful as possible in identifying and measuring errors.  In particular, we have extensively used Monte Carlo simulations to model both systematic and random errors in our measurements.  Although we were unable to independently check the Holtzman (1995b) zero points to a precision of $<$0.1 magnitudes through our ground-based photometry, our measurements of the abundances and relative ages are differential, and so should not be affected by errors in the zero points.  The reddenings and distance moduli that we derive, however, {\it are} subject to unkown zero point errors in the photometry.  

Adopting the CMD-based abundances, the possibility that the old LMC globular clusters are younger than the oldest Milky Way globular clusters by 2-3 Gyr (Da Costa 1993) appears not to be the case for the clusters studied here.  This result is clear from the relative ages derived from the comparison of the clusters with Milky Way fiducials, the results of which are shown in Table 6.  NGC 1754, NGC 1898, NGC 2005, and NGC 2019, which we find have abundances most closely matching that of the classical old Galactic globular cluster M5, are all of similar age or older than M5.  NGC 1835, which we find has an abundance similar to M3, is $\sim$2 Gyr older.  As LDZ consider M3 to be $\sim$2 Gyr younger than the oldest halo clusters, NGC 1835 appears to have an age similar to the oldest Milky Way clusters.

Interpreting our data through the HB models of LDZ, we paint a similar picture.  As shown in Fig. 17a, the clusters NGC 1754, NGC 1898, NGC 2005, and NGC 2019 all fall on the HB evolutionary track crossing through the oldest Milky Way halo clusters.  Comparing our data with the HB data of Walker (1992b) for the outer LMC old globular clusters, we find that the HB models suggest that the earliest episode of cluster formation in the LMC spanned 2-3 Gyr.  However, we find some problems with the ages implied by the HB models.  The age of NGC 1835 suggested by the HB models is inconsistent with its age relative to M3.  In addition, the age of NGC 1754 relative to M5 is slightly older than suggested by the HB models.  These inconsistencies may indicate a difficulty with the interpretation that metallicity and age are the only parameters affecting HB morphology, as has been suggested by many authors (e.g. Buonanno et al. 1997, Catelan et al. 1997, Sweigart 1997).  Given the currently large errors in the abundance measurements, however, we do not claim on the basis of our data that this interpretation needs to be modified.

The ages we derive from $\Delta V{\rm ^{TO}_{HB}}$ are marginally consistent with the picture that the LMC clusters studied here are as old as the oldest Milky Way clusters.  Our measured values of $\Delta V{\rm ^{TO}_{HB}}$ imply an average age of 15.3 $\pm$ 1.5 Gyr, which is lower but within $\sim1.5\sigma$ of the average age of the old halo quoted by Chaboyer et al. (1996), 17.8 $\pm$ 0.4 Gyr.

Because of the metallicity dependence of the age indicators discussed in Section 6.2, our conclusions based on the cluster ages depend critically on the CMD-based abundances being correct.  If we adopt the O91 spectroscopic abundances, for instance, our conclusions change considerably.  For example, Fig. 17b shows the comparison of the LDZ HB evolutionary tracks with our cluster data using the O91 abundances.  The HB tracks imply, in this case, that the LMC clusters are indeed $\sim$2 Gyr younger than the old Milky Way halo.  However, we are then unable to build a consistent picture with the relative ages derived from the comparison to Milky Way clusters.  Coupled with the better internal consistency weakly implied by the distance moduli (Figs. 18a-d), we continue to prefer the CMD-based abundances for these LMC clusters.  Nevertheless, it would clearly be extremely valuable to have high-resolution spectroscopic abundances of several stars in each cluster available.

In contrast to the globular clusters of the Milky Way halo, the old LMC clusters are not as clearly part of a distinct halo (O91; Schommer et al. 1992).  Our results imply that at the time the Milky Way formed its first globular clusters, the LMC may have already collapsed to a disk and started forming clusters.  Why there appears to be no clear halo component in the LMC (Olszewski, Suntzeff, \& Mateo 1996) remains an open question.  However, by establishing that the oldest clusters in the LMC are truly as old as the Milky Way Galaxy, we have taken an important step towards understanding the formation of the LMC.

\section*{Acknowledgements}
We are indebted to Michael Bolte and Jennifer Johnson for providing us with the 
unpublished $(V-I),V$ fiducial sequences of M3, M5, M13, and M92 and to Georgi Mandushev for the unpublished sequence of M55.  We are grateful to Don VandenBerg for his invaluable help with the cluster ages and to Carlton Pryor for his help with the King model fits.  We thank Eric Deutsch for his modifications to DoPHOT which made it easier to use.  The anonymous referee is thanked for helpful comments which improved the manuscript, as is George Wallerstein for his reading and comments.  This work was supported by STScI grant GO05916.

\end{document}